\documentclass{IEEEtran}
\usepackage{mathpazo,times,amssymb,amsmath,amsfonts,float,nicefrac,subfigure,float,accents,color}
\usepackage{euscript,graphics}
\usepackage[dvips]{epsfig}

\newtheorem{theorem}{Theorem}[section]
\newtheorem{lemma}[theorem]{Lemma}
\newtheorem{proposition}[theorem]{Proposition}
\newtheorem{corollary}[theorem]{Corollary}

\newcommand{\remove}[1]{}

\newtheorem{cnstr}{Construction}

\newtheorem{xmpl}{Example}

\newcommand{\ff}{\mathbb{F}}

\newcommand{\ceilenv}[1]{\left\lceil #1 \right\rceil}
\newcommand{\floorenv}[1]{\left\lfloor #1 \right\rfloor}

\DeclareMathOperator{\ord}{ord}

\newcommand\nc\newcommand
\nc\bfa{{\boldsymbol a}}\nc\bfA{{\bf A}}\nc\cA{{\mathcal A}}
\nc\bfb{{\boldsymbol b}}\nc\bfB{{\bf B}}\nc\cB{{\mathcal B}}
\nc\bfc{{\boldsymbol c}}\nc\bfC{{\bf C}}\nc\cC{{\mathcal C}}
\nc\bfd{{\boldsymbol d}}\nc\bfD{{\bf D}}\nc\cD{{\mathcal D}}
\nc\bfe{{\boldsymbol e}}\nc\bfE{{\bf E}}\nc\cE{{\mathcal E}}
\nc\bff{{\boldsymbol f}}\nc\bfF{{\bf F}}\nc\cF{{\mathcal F}}
\nc\bfg{{\boldsymbol g}}\nc\bfG{{\bf G}}\nc\cG{{\mathcal G}}
\nc\bfh{{\boldsymbol h}}\nc\bfH{{\bf H}}\nc\cH{{\mathcal H}}
\nc\bfi{{\boldsymbol i}}\nc\bfI{{\bf I}}\nc\cI{{\mathcal I}}
\nc\bfj{{\boldsymbol j}}\nc\bfJ{{\bf J}}\nc\cJ{{\mathcal J}}
\nc\bfk{{\boldsymbol k}}\nc\bfK{{\bf K}}\nc\cK{{\mathcal K}}
\nc\bfl{{\boldsymbol l}}\nc\bfL{{\bf L}}\nc\cL{{\mathcal L}}
\nc\bfm{{\boldsymbol m}}\nc\bfM{{\bf M}}\nc\cM{{\mathcal M}}
\nc\bfn{{\boldsymbol n}}\nc\bfN{{\bf N}}\nc\cN{{\mathcal N}}
\nc\bfo{{\boldsymbol o}}\nc\bfO{{\bf O}}\nc\cO{{\mathcal O}}
\nc\bfp{{\boldsymbol p}}\nc\bfP{{\bf P}}\nc\cP{{\mathcal P}}
\nc\bfq{{\boldsymbol q}}\nc\bfQ{{\bf Q}}\nc\cQ{{\mathcal Q}}
\nc\bfr{{\boldsymbol r}}\nc\bfR{{\bf R}}\nc\cR{{\mathcal R}}
\nc\bfs{{\boldsymbol s}}\nc\bfS{{\bf S}}\nc\cS{{\mathcal S}}
\nc\bft{{\boldsymbol t}}\nc\bfT{{\bf T}}\nc\cT{{\mathcal T}}
\nc\bfu{{\boldsymbol u}}\nc\bfU{{\bf U}}\nc\cU{{\mathcal U}}
\nc\bfv{{\boldsymbol v}}\nc\bfV{{\bf V}}\nc\cV{{\mathcal V}}
\nc\bfw{{\boldsymbol w}}\nc\bfW{{\bf W}}\nc\cW{{\mathcal W}}
\nc\bfx{{\boldsymbol x}}\nc\bfX{{\bf X}}\nc\cX{{\mathcal X}}
\nc\bfy{{\boldsymbol y}}\nc\bfY{{\bf Y}}\nc\cY{{\mathcal Y}}
\nc\bfz{{\boldsymbol z}}\nc\bfZ{{\bf Z}}\nc\cZ{{\mathcal Z}}

\newcommand{\bfit}{\bfseries\itshape}

\DeclareSymbolFont{bbold}{U}{bbold}{m}{n}
\DeclareSymbolFontAlphabet{\mathbbold}{bbold}

\title{\Large\bf A family of optimal locally recoverable codes}

\author{Itzhak Tamo, {\em Member}, IEEE, and Alexander Barg, {\em Fellow}, IEEE

\thanks{
\noindent Manuscript received October 10, 2013; revised February 6, 2014; accepted
April 18, 2014.

The authors' emails are \{zactamo,alexanderbarg\}@gmail.com.
Provisional US patent application 61/884,768 filed.

Communicated by H. Pfister, Associate Editor for Coding Theory.

Digital Object Identifier 10.1109/TIT.2014.2321280 
 
   }
}

\begin{document}\maketitle
\thispagestyle{empty}

\begin{abstract} A code  over a finite alphabet is called locally recoverable (LRC) if every symbol in the encoding is a function of a small number (at most $r$) other symbols.
 We present a family of LRC codes that attain the maximum possible value
of the distance for a given locality parameter and code cardinality. 
The codewords are obtained as evaluations of specially constructed polynomials over a finite field, 
and reduce to a Reed-Solomon code if the locality parameter $r$ is set to be equal to the code dimension. 
The size of the code alphabet for most parameters is only slightly greater than the code length.
The recovery procedure is performed by polynomial interpolation over $r$ points.
We also construct codes with several disjoint recovering sets for every symbol. This construction enables the system to conduct several independent and simultaneous recovery processes of a specific symbol by accessing different parts of the codeword. This property enables high availability of  frequently accessed data (``hot data'').
\end{abstract}

%
%
\section{Introduction}
Distributed and cloud storage systems have reached such a massive scale that recovery from several failures is now part of
regular operation of the system rather than a rare exception. In addition, storage systems have to provide high data availability to ensure high performance. In order to address these requirements, redundancy and data encoding must be introduced into the system. The simplest and most widespread technique used for data recovery is replication, under which several copies of each
data fragment are written to distinct physical storage nodes.
However, this solution entails large storage overhead and has therefore become inadequate for
modern  systems supporting the ``Big Data'' environment. 
Therefore, more advanced coding techniques that provide comparable resiliency against failures 
at the expense of a significantly smaller storage overhead, are implemented. 
For example, Facebook uses the $(14,10)$ Reed-Solomon code, which requires only $40\%$ overhead compared to the $200\%$ overhead associated with threefold replication. 

Although today's storage systems are resilient to several concurrent node failures, in order to provide enough data reliability, by far the most common scenario is a failure of a single node. Hence, a storage system should be designed to efficiently repair such scenarios. The repair efficiency of a single node failure in the system can be quantified under different metrics, where each metric is relevant for different storage systems and applications.   
More precisely, a large body of existing work has considered the repair problem under three metrics:
i) the number of bits communicated in the network, i.e., the repair-bandwidth \cite{dimakis2010network,rashmi2011optimal,suh2011exact,Zigzag-code,cadambe2011optimal,papailiopoulos2011repair},
ii), the number of bits read, the disk-I/O \cite{khan2011search,Zigzag-code}, and
iii), repair locality, i.e., the number of nodes that participate in the repair process \cite{gopalan2011locality,oggier2011self,papailiopoulos2012simple,My-paper,Natalia}.
The fundamental limits of these metrics are yet to be fully understood.
In this work, we focus on the last of these metrics, namely the repair locality. 

More formally, a Locally Recoverable Code (LRC code) of length $n$ is a code that produces an $n$-symbol codeword from $k$ information symbols and, for \emph{any} symbol  of the codeword, there exist at most $r$ other symbols such that the value of the symbol can be recovered from them. We refer to such a code as an $(n,k,r)$ LRC code. 
For LRC codes, if a symbol is lost due to a node failure, its value can be recovered by accessing the value of at most $r$ other symbols. For example, a code of length $2k$ in which each coordinate is repeated twice, is an LRC code with locality $r=1.$ 
Generally the locality parameter satisfies $1\le r\le k$ because the entire codeword can be found by accessing $k$ symbols other
than the erased symbol.
Another example is given by $(n,k)$ maximum distance separable, or MDS codes. In this case the locality is $r=k$, and not less than that, which is the largest possible value.
Observe that MDS codes can recover the largest possible number of erased symbols among all $(n,k)$ codes, but
they are far from optimal in terms of locality, i.e., for correcting a single symbol erasure. 
Yet another simple example is provided by regular LDPC codes with $r+1$ nonzeros in every check equation, 
meaning that any single symbol of the codeword is a linear combination of some other $r$ symbols.

Codes that have good locality properties were
initially studied in \cite{han2007reliable}, \cite{huang2007pyramid}, although the second of these papers
considered a slightly different definition of locality, under which a code is said to have {\em information locality} $r$ 
if the value of any of its information symbols can be recovered by accessing at most $r$ other codeword symbols. 
Codes with information locality property were also studied in  \cite{gopalan2011locality,yekhanin-locality-nonlinear}.   
A natural question to ask is as follows: given an $(n,k,r)$ LRC code $\cC,$ what is the best possible minimum distance $d(\cC)$? 
A bound on $d(\cC)$ as a function of $n,k$ and $r$ was proved in \cite{gopalan2011locality} by extending the arguments
in the proof of the classical Singleton bound on codes (see Theorem \ref{thm:S} below).
Using a probabilistic argument, \cite{gopalan2011locality} showed that this bound is tight over a large enough finite field. Therefore, an $(n,k,r)$ LRC code that achieves the bound of \cite{gopalan2011locality}  with equality is called an optimal LRC code. 
The Singleton-type bound of \cite{gopalan2011locality} does not take into account the cardinality of the code alphabet $q$.
Augmenting this result, a recent work 
\cite{2013arXiv1308.3200C} established a bound on the distance of LRC codes that depends on $q$, sometimes 
yielding better results. 
Another perspective of the limits for LRC codes was addressed in \cite{Maz13} which showed that locality cannot be too small if the codes are required to attain capacity of, say, the binary symmetric channel. We note that locality enables one to recover from a single
failure with only $r$ reads and thus offers a significant speedup in the most common scenario.

There are two constructions of optimal LRC codes known in the literature. Namely, \cite{Natalia} proposed a two-level 
construction based on the well-known Gabidulin codes combined with a single parity-check $(r+1,r)$ code. 
Another construction ~\cite{My-paper} used two layers of MDS codes,
a Reed-Solomon code and a special $(r+1,r)$ MDS code.
A common shortcoming of these constructions relates to the size of the code alphabet which in both papers is an exponential 
function of the code length, complicating the implementation. 
The only known constructions of
optimal LRC codes over an alphabet of size comparable to code's length are for locality $r=1,k$, and recently paper \cite{prakash2012optimal} constructed such a code for a specific value of the length $n=\lceil \frac{k}{r}\rceil
(r+1)$.
In this paper we overcome this shortcoming,
presenting a natural generalization of the Reed-Solomon construction which relies on the alphabet of cardinality comparable to the code length $n$. Our construction can also be viewed in the framework of codes constructed using the Chinese Remainder Theorem; see Sect.~\ref{sect:RRC}.

Recently \cite{Lluis2013} constructed LRC codes with several disjoint repair alternatives using partial geometries.
\cite{Rashmi2013piggy} presented a new framework for designing distributed storage codes that are efficient in data
read and download required during repair, and \cite{kamath2012codes} presented codes that combine two metrics related
to storage, namely codes with local recovery that at the same time seek to minimize the repair bandwidth during repair of a failed node.

A related  locality property, introduced in \cite{Blaum-Partial-MDS-Codes,Maximally-Recoverable-Codes-with-Locality}, 
is
called maximally recoverable codes. Symbols in such codes can be grouped into disjoint sets of size $r+1$ that form a simple parity check code. Moreover, puncturing each codeword on
one coordinate from each group yields an MDS code. Hence the value of each symbol in such codes can be recovered by a simple parity check sum of $r$ other symbols.

\vspace*{.05in}
{\em Overview of the paper:} 
The main construction of optimal $(n,k,r)$ LRC codes over the finite field $\ff_q, q\ge n$ is presented in Section \ref{sect:construction}.
There are several versions of the construction that are discussed in detail, together with some examples of short optimal LRC codes. We 
also observe that the encoding can be made systematic, which may be beneficial in implementations. In Section \ref{sect:multiple} we
give two constructions of LRC codes with multiple disjoint recovering sets for each symbol, which enables simultaneous recovery from
different portions of the encoding. In Section~\ref{sect:generalizations} we discuss several extensions of the main construction, 
in particular, pointing out that the simplifying assumptions made earlier in the paper can be removed with only small changes in the 
resulting codes.

\vspace*{.05in}
Throughout the paper, $\cC$ denotes a code over a finite field $\ff_q$. The triple of parameters $(n,k,r)$ refers to a code of length $n,$
cardinality $q^k$ and locality $r$. The finite field is also denoted by $\ff$ if its cardinality is understood or does not matter.
We also use the notation $[n]:=\{1,\dots,n\}$. A {\em restriction} $\cC_I$ of the code $\cC$ to a subset of coordinates $I\subset[n]$
is the code obtained by removing from each vector the coordinates outside $I.$

\section{Preliminaries on LRC codes}\label{sect:prelim}
We say that a code $\cC\subset \ff_q^n$ has locality $r$ if every symbol of the codeword $x\in \cC$ 
can be recovered from a subset of $r$ other symbols of $x$ (i.e., is a function of some other $r$ symbols $x_{i_1},x_{i_2},\dots,x_{i_r}$).
In other words, this means that, given $x\in \cC, i\in [n],$ there exists a subset of coordinates 
$I_i\subset [n]\backslash i, |I_i|\le r$ such that
the restriction of $\cC$ to the coordinates in $I_i$ enables one to find the value of $x_i.$ The subset 
$I_i$ is called a {\em recovering set} for the symbol $x_i$. 

The formal definition is as follows. Given $a\in \ff_q$ consider the sets of codewords
   $$
   \cC(i,a)=\{x\in \cC: x_i=a\},\quad i\in[n].
   $$
    The code $\cC$ is said to have {\em locality} $r$ if for every $i\in [n]$ there exists a subset $I_i\subset [n]\backslash i, |I_i|\le r$
    such that the restrictions of the sets $\cC(i,a)$ to
the coordinates in $I_i$ for different $a$ are disjoint: $$\cC_{I_i}(i,a)\cap \cC_{I_i}(i,a')=\emptyset,\quad  a\ne a'.$$

The code $\cC_{I_i\cup \{i\}}$ is called a {\em local code} of the code $\cC$. In the constructions
of LRC codes presented in the literature the set of coordinates of the $(n,k,r)$ LRC code is usually partitioned into 
$(r+1,r)$ local MDS codes that define the recovering sets of the symbols.

Two desirable features of codes are large minimum distance and high rate.
We begin with two bounds on these parameters of an LRC code. 

The proof of the following theorem is given in the appendix.

\begin{theorem} \label{thm:S}
Let $\cC$ be an $(n,k,r)$ LRC code of cardinality $q^k$ over an alphabet of size $q$, then: \\
   The rate of $\cC$ satisfies
				\begin{equation}
							\frac{k}{n}\leq  \frac{r}{r+1}.
							\label{prop546}
				\end{equation}
				The minimum distance of $\cC$ 
				satisfies 
   			\begin{equation}
				d\leq n-k-\ceilenv{ \frac{k}{r}}+2.				 
				\label{eq:erer}
     		\end{equation}
A code that achieves the bound on the distance with equality will be called an {\em optimal LRC code}.
\end{theorem}
\noindent{\em Remark:} The bound on the distance is due to \cite{gopalan2011locality,LRC_Dimitiris}, where it appears with a different proof.

It is clear that in any code, each symbol has locality at most $k$, so $r$ always satisfies $1\leq r\leq k.$
Upon letting $r=k,$ \eqref{eq:erer} becomes
the well-known Singleton bound, 
   \begin{equation}\label{eq:Singleton}
   d\le n-k+1,
   \end{equation}
so optimal LRC codes with $r=k$ are precisely MDS codes, e.g. the Reed-Solomon codes.
On the other hand, if $r=1,$ the bound \eqref{eq:erer} becomes
  $$
  d\leq n-2k+2=2\Big(\frac{n}{2}-k+1\Big).
  $$
Replicating each symbol twice in an $(n/2,k)$ MDS code, we obtain an optimal LRC code with locality $r=1$.\

\section{Code Construction}\label{sect:construction}
In this section we construct optimal linear $(n,k,r)$ LRC codes over a finite field alphabet of size $q,$ where $q$ is a prime power
greater or equal to $n$.
In the first version of the construction we assume that  $k$ is divisible by $r$ (this restriction will be removed later in this section). 
Throughout this section we also assume that $n$ is divisible by $r+1$ (this restriction can also be lifted, see Sect.~\ref{sect:arbitrary}).

\subsection{General construction} We begin with a general method of constructing linear codes with the locality property. 
Later we will show that some of these codes have optimal minimum distance. 
The codes are constructed as evaluations of polynomials, in line with many other algebraic code constructions.
Unlike the classical Reed-Solomon codes, the new codes will be evaluated at a specially chosen set of points of the field $\ff_q, q\ge n.$ 
A key ingredient of the construction is a polynomial $g(x)\in\ff_q[x]$ that satisfies the following conditions:
\begin{enumerate}
	\item The degree of $g$ is $r+1$,
	\item There exists a partition $\cA=\{A_1,...,A_{\frac{n}{r+1}}\}$ of a set $A\subseteq \ff_q$ of size $n$ into sets of size $r+1$, such that $g$ is constant on each set $A_i$ in the partition. Namely for all $i=1,\dots,n/(r+1),$ and any $\alpha,\beta\in A_i$,
	$$g(\alpha)=g(\beta).$$  
\end{enumerate}
A polynomial that satisfies these conditions will be called {\em good}. The code construction presented below
relies on the existence of good polynomials. 

\begin{cnstr} \label{cnstr3} {\em $($$(n,k,r)$ \emph{LRC codes}$)$ Let $n\le q$ be the target code length. 
Let $A\subset \ff_q, |A|=n$ and let $g(x)$ be a good polynomial for the partition $\cA$ of the set $A$.
To find the codeword for a message vector 
$a\in\ff_q^k$ write it as $a=(a_{ij},i=0,\dots,r-1;  j=0,...,\frac{k}{r}-1).$
Define the encoding polynomial 
\begin{equation}
f_a(x)=\sum_{i=0}^{r-1}f_i(x)x^i,
\label{eq:97979}
\end{equation}
where 
\begin{equation}
f_i(x)=\sum_{j=0}^{\frac{k}{r}-1}a_{ij}g(x)^j, \quad i=0,...,r-1
\label{eq:jnjn}
\end{equation}
(we call the $f_i$'s the coefficient polynomials). The codeword for $a$ is found as the evaluation vector of $f_a$ 
at all the points of $A$. In other words, the $(n,k,r)$ LRC code $\cC$ is defined as the set of $n$-dimensional vectors
  \begin{equation}\label{eq:code1}
    \cC=\{(f_a(\alpha),\alpha\in A): a\in \ff_q^k\}.
  \end{equation}
  We call the elements of the set $A$ {\em locations} and the elements of the vector
$(f_a(\alpha))$ {\em symbols} of the codeword.}
\end{cnstr}

 The local recovery is accomplished as follows.

\vspace*{.05in}
{\bfit Recovery of the erased symbol:} Suppose that the erased symbol corresponds to the location $\alpha\in A_j,$
where $A_j$ is one of the sets in the partition $\cA.$ Let $(c_\beta, \beta\in A_j\backslash\alpha)$ denote the
remaining $r$ symbols in the locations of the set $A_j$. 
To find the value $c_\alpha=f_a(\alpha)$, find the unique polynomial $\delta(x)$ of degree less than $r$ such that
$\delta(\beta)=c_\beta$ for all $\beta\in A_j\backslash\alpha,$ i.e., 
  \begin{equation}\label{eq:dec}
\delta(x)=\sum_{\beta\in A_j\backslash\alpha} c_\beta\prod_{\beta'\in A_j\backslash\{\alpha,\beta\}} \frac{x-\beta'}{\beta-\beta'}
  \end{equation}
and set $c_\alpha=\delta(\alpha).$  We call
$\delta(x)$ the {\em decoding polynomial} for the symbol $c_\alpha.$ Thus, to find one erased symbol, we need to perform polynomial interpolation from $r$ known symbols in its recovery set. This recovery procedure underlies all the constructions in this paper.

\vspace*{.05in}

In the next theorem we prove that the codes constructed above are optimal with respect to the bound \eqref{eq:erer}, and
justify the validity of the recovery procedure.
\begin{theorem} \label{thm1.1} 
The linear code $\cC$ defined in \eqref{eq:code1} has dimension $k$ and is an 
optimal $(n,k,r)$ LRC code, namely its minimum distance meets the bound \eqref{eq:erer} with equality. 
\remove{  \begin{equation}\label{eq:erer}
  d=n-k-\frac{k}{r}+2.
  \end{equation}}
\end{theorem}
\begin{IEEEproof} 
Note that for $i=0,...,r-1;j=0,...,\frac{k}{r}-1$ the $k$ polynomials $g(x)^jx^i$ all are of distinct degrees, and therefore
are linearly independent over $\ff.$  In other words, the mapping $a\mapsto f_a(x)$ is injective. By \eqref{eq:97979}, \eqref{eq:jnjn} the degree of the polynomial $f_a(x)$ is at most
  \begin{equation*}
    \Big(\frac kr-1\Big)(r+1)+r-1= k+\frac{k}{r} -2\leq n-2,
  \end{equation*}
where the last inequality follows from \eqref{prop546}. This means that two distinct encoding
polynomials $f_a$ and $f_b$ give rise to two distinct codevectors, so the dimension of the code is $k$. Since the encoding is linear, the distance satisfies
  $$
    d(\cC)\ge n-\max_{f_a, a\in \ff_q^k}\deg(f_a)=n-k-\frac{k}{r}+2
  $$
which together with \eqref{eq:erer} completes the proof of distance optimality.

Let us prove the locality property. Let $A_j$ be a member of the partition $\cA$ and assume that the
lost symbol of the codeword equals $c_\alpha=f_a(\alpha),$ where $\alpha\in A_j$ is a field element. 
 Define the {\em decoding polynomial}
\begin{equation}
\partial(x)=\sum_{i=0}^{r-1}f_i(\alpha)x^i,
\label{eq:decoding2}
\end{equation}
where the $f_i(x)$ are the coefficient polynomials \eqref{eq:jnjn}. We will show that $\partial(x)$ is the same polynomial as $\delta(x)$ defined in \eqref{eq:dec}. Each $f_i(x)$ is a linear combination of powers of $g$, therefore it is also constant on the set $A_j$, 
i.e., for any $\beta \in A_j$ and any coefficient polynomial $f_i, i=1,\dots,r-1$ 
\begin{equation}
f_i(\beta)=f_i(\alpha).
\label{eq:11111}
\end{equation}
Hence by \eqref{eq:decoding2} and \eqref{eq:11111}, for any $\beta$ in $A_j$
   $$
\partial(\beta)=\sum_{i=0}^{r-1}f_i(\alpha)\beta^i=\sum_{i=0}^{r-1}f_i(\beta)\beta^i=f_a(\beta).
$$ 
In other words, the values of the encoding polynomial $f_a(x)$ and the decoding polynomial $\partial(x)$ on the locations of $A_j$ coincide. 
Since $\partial(x)$ is of degree at most $r-1$, 
it can be interpolated from  the $r$ symbols $c_\beta,\beta\in A_j\backslash \alpha,$ cf.~Eq.~\eqref{eq:dec}.
Once $\partial(x)$ is computed, we find the lost symbol as $\partial(\alpha).$ 
To conclude, the lost symbol $c_\alpha$
can be recovered by accessing $r$ other symbols of the codeword.
\end{IEEEproof}

As a consequence of this proof, we note that the polynomial $\delta(x)$ satisfies the condition $\delta(\alpha)=f_a(\alpha)$ 
for all $\alpha\in A_j,$ i.e., it is determined by the index $j$ of the recovering set $A_j$. In other words, the decoding polynomial $\delta(x)$ is the same for any two symbols $\alpha_1,\alpha_2 \in A_j$. 

\begin{xmpl}{\rm In this example we construct an optimal $(n=9,k=4,r=2)$ LRC code over the field $\ff_q.$
Since we need $9$ distinct evaluation points of the field, we must choose $q\ge 9$. 
We define the code $\cC$ over $\ff_{13}.$ 

The difficulty of using Construction \ref{cnstr3} is in constructing a 
good polynomial $g$ of degree $r+1=3$ that is constant on $3$ disjoint sets of  size $3$. 
In this example we offer little motivation in constructing $g(x)$ but later we will give a systematic way
of constructing them. 

Let the partition $\cA$ be as follows:
    $$
\cA=\{A_1=\{1,3,9\},A_2=\{2,6,5\},A_3=\{4,12,10\}\},
    $$ 
and note that the polynomial $g(x)=x^3$ is constant on the sets $A_i.$ 
Let $a=(a_{0,0},a_{0,1},a_{1,0},a_{1,1})$ be the information vector of length $k=4$ over $\ff_{13}$ 
and define the encoding polynomial by \eqref{eq:97979}, \eqref{eq:jnjn}
    \begin{align*} 
f_a(x)&=(a_{0,0}+a_{0,1}g(x))+x(a_{1,0}+a_{1,1}g(x))\\
   &=(a_{0,0}+a_{0,1}x^3)+x(a_{1,0}+a_{1,1}x^3)\\&=a_{0,0}+a_{1,0}x+a_{0,1}x^3+a_{1,1}x^4.
   \end{align*}
The codeword $c$ that corresponds to $a$ 
is found as the evaluation of the polynomial $f_a$ at all the points of the sets of the partition $\cA$: ${c=(f_a(\alpha),\,\alpha\in \cup_{i=1}^3 A_i)}.$ Since $\deg f_a\le 4,$ the minimum distance is at least $5$,
and so $d=5$ by \eqref{eq:erer}. 
For instance, assume that $a=(1,1,1,1),$ then the codeword is found to be 
\begin{gather*}
(f_a(1),f_a(3),f_a(9),f_a(2),f_a(6),f_a(5),f_a(4),f_a(12),f_a(10))\\=(4,8,7,1,11,2,0,0,0).
\end{gather*}

Suppose that the value $f_a(1)$ is erased. By our construction, it can be recovered
by accessing 2 other codeword symbols, namely, the symbols at locations corresponding to 3 and 9. Using \eqref{eq:dec} we find $\delta(x)=2x+2$ and compute $\delta(1)=4,$ which is the required value.
}
\label{xmpl1}
\end{xmpl}

{\bf Remarks:}
\begin{enumerate}
	\item Construction \ref{cnstr3} is a direct extension of the classical \emph{Reed-Solomon}  codes in that both are
evaluations of some polynomials defined by the message vector. Our construction also reduces to Reed-Solomon
codes if $r$ is taken to be $k$. Note that if $r=k$ then each coefficient polynomial  \eqref{eq:jnjn} is a constant, and
therefore the code construction does not require a good polynomial.
For the same reason, the set $A$ for 
RS codes can be an arbitrary subset of $\ff_q$, while the locality condition for $r<k$ imposes a restriction on the choice
of the locations.

\item Note that if the coordinates of the vector $a$ are indexed as $a=(a_0,...,a_{k-1})$ 
then the encoding polynomial in \eqref{eq:97979} can be also written as 
    \begin{equation}\label{eq:clever}
    f_a(x)=\sum_{\substack{m=0\\m\neq r\,\text{mod}(r+1)}}^{k+\frac{k}{r}-2}a_mg(x)^{\lfloor \frac{m}{r+1}\rfloor}x^{m\,\text{mod} (r+1)}.
    \end{equation}
To see this, put in \eqref{eq:jnjn} $a_m=a_{i+j(r+1)},i=0,\dots,r-1;j=0,\dots,\frac kr-1,$ and observe that 
$\lfloor\frac{k+\frac{k}{r}-2}{r+1}\rfloor=\frac kr-1,$
and that there are $k/r-1$ numbers in the set $\{0,1,\dots,k+(k/r)-2\}$ equal to $r$ modulo $r+1.$
	\item In Construction \ref{cnstr3} we assumed that $r$ divides $k$; however, this constraint can
be easily lifted.
Indeed, suppose that $r$ does not divide $k$ and define the coefficient polynomial $f_i$ in \eqref{eq:jnjn} as follows: 
       $$
   f_i(x)=\sum_{j=0}^{s(k,r,i)} a_{ij} g(x)^j, \quad i=0,1,\dots, r-1,
      $$
where
    $$
   s(k,r,i)=\begin{cases}
\lfloor \frac{k}{r}\rfloor & i< k\,\text{mod}\, r \\\,
\lfloor \frac{k}{r}\rfloor-1 & i\geq k\,\text{mod}\, r.
\end{cases}
    $$
It is easy to see that the $r$ coefficient polynomials are defined by the $k$ information symbols, and the resulting encoding polynomial $f_a$ has degree at most $k+\lceil k/r\rceil-2$. The remaining parts of the 
construction are unchanged.
\end{enumerate}

\subsection{Constructing optimal LRC codes using algebraic structure of the  field}
The main component of Construction \ref{cnstr3} is finding a good polynomial $g(x)$ together with the corresponding
partition of the subset $A$ of the field. In this section we show how to construct $g(x)$
 using the multiplicative and additive groups of $\ff_q$. 

The multiplicative group $\ff_q^\ast$ is cyclic, and the additive group $\ff_q^+$ is isomorphic to 
a direct product of $l$ copies of the additive group $\mathbb{Z}_p^+$, where $q=p^l$ and $p$ is the characteristic of the field. 
The following obvious proposition constructs a good polynomial from any subgroup of $\ff_q^\ast$ or $\ff_q^+.$
\begin{proposition}
Let $H$ be a subgroup of $\ff_q^\ast$ or $\ff_q^+.$ The annihilator polynomial of the subgroup
\begin{equation}
g(x)=\prod_{h\in H}(x-h)
\label{eq:nknk}
\end{equation}
is constant on each coset of $H$. 
\label{lem:good}
\end{proposition}
\begin{IEEEproof}
Assume that $H$ is a multiplicative subgroup and let $a,a\overline{h}$ be two elements of the coset $aH$, where $\overline{h}\in H$, then
\begin{align}
g(a\overline{h})=\prod_{h\in H}(a\overline{h}-h)=&\overline{h}^{|H|}\prod_{h\in H}(a-h\overline{h}^{-1})\nonumber\\
=&\prod_{h\in H}(a-h)\nonumber\\
=&g(a)\nonumber.
\end{align}
The proof for additive subgroups is completely analogous.
\end{IEEEproof}

{\em Remark:} If $H$ is a multiplicative subgroup of ${\mathbb F}_q^\ast$, then $g(x)$ in \eqref{eq:nknk} can be written as
 $g(x)=x^{|H|}-1.$ Equivalently, we can take $g(x)=x^{|H|}.$
 
\remove{First assume that $H$ is an additive subgroup and let $a,a+\overline{h}$ be two elements of the coset $a+H$, where $\overline{h}\in H$, then
\begin{align*}
g(a+\overline{h})=&\prod_{h\in H}(a+\overline{h}-h)\\
=&\prod_{h\in H}(a-h)\\
=&g(a).
\end{align*}
}

Thus annihilators of subgroups form a class of good polynomials that can be used to construct optimal codes.
The partition $\cA$ is a union of cosets of $H$, so the code length $n$ can be any multiple of $r+1$ satisfying
$n\le q-1$ (or $n\le q$ in the case of the additive group). 
Since the size of the subgroup divides the size of the group we get that $q\,\text mod\,(r+1)$ is $1$ (or $0$).

The parameters of LRC codes constructed using subgroups are naturally restricted by the possible size of the 
subgroups. 
Note that Example \ref{xmpl1} is constructed using the multiplicative subgroup $H=\{1,3,9\}$ of the field $\ff_{13},$ 
and the annihilator is $g(x)=x^3-1$. In the example we used another good polynomial, $g(x)=x^3.$

\begin{xmpl}{\rm
In this example we construct an optimal $(12,6,3)$  LRC code with minimum distance $d=6$ over $\ff_{13}.$ 
Note that  
$5$ is an $(r+1)=4$-th root of unity modulo 13, therefore
 the polynomial $g(x)=x^4$ is constant on the cosets of the cyclic group $H=\{1,5,12,8\}$
generated by $5.$ Note that the polynomial $g$ constructed in Proposition \ref{lem:good} is in fact $g(x)=x^4-1$, while we use the 
polynomial $g(x)=x^4$. 
Since the polynomials $1,x^4-1$ span the same subspace as the polynomials $1,x^4,$ the resulting
codes are equivalent. 

 The group $H$ gives rise to the partition of $\ff_{13}^\ast$
   \begin{multline*}
   \cA=\Big\{A_1=\{1,5,12,8\},\\A_2=\{2,10,11,3\},A_3=\{4,7,9,6\}\Big\}.
   \end{multline*}
For the information vector $(a_0,a_1,a_2,a_4,a_5,a_6)$ define the encoding polynomial \eqref{eq:clever}
   $$
f_a(x)=\sum_{\substack{i=0\\ i\neq 3}}^6a_ix^i=f_0(x)+f_1(x)x+f_2(x)x^2
    $$
with  coefficient polynomials equal to
   $$
f_0(x)=a_0+a_4x^4,\; f_1(x)=a_1+a_5x^4,\; f_2(x)=a_2+a_6x^4. 
   $$ 
The corresponding codeword is obtained by evaluating $f_a(x)$ for all the points $x\in \ff_{13}^\ast.$

}
\end{xmpl}
\begin{xmpl}{\rm
In this example we construct an optimal LRC code using the additive group of the field. 
Let $\alpha$ be a primitive element of the field $\ff_{2^4}$ and take the additive subgroup
$H=\{x+y\alpha:x,y\in \ff_2\}.$ The polynomial $g(x)$ in \eqref{eq:nknk} equals 
    \begin{align*}
g(x)&=x(x+1)(x+\alpha)(x+\alpha+1)\\&
=x^4+(\alpha^2+\alpha+1) x^2+(\alpha^2+\alpha) x.
   \end{align*}

We will construct an optimal $(12,6,3)$ LRC code with distance $d=6$. For $i=0,1,2$ define the coefficient polynomials 
    $$
f_i(x)=a_{i,0}+a_{i,1}g(x),
    $$ 
using the information vector $a=(a_{i,j})$ and $i=0,1,2$, $j=0,1$. 
The subgroup $H$ is of order $4$, hence in order to have $12$ evaluation points, we choose any 
$3$ cosets of $H$ out of its $4$ cosets, and evaluate the encoding polynomial 
    $$
f_a(x)=f_2(x)x^2+f_1(x)x +f_0(x)
    $$ 
at the elements of these cosets.
Theorem \ref{thm1.1} implies that the resulting code has the claimed properties.
Comparing this code with a $(12,6)$ MDS code, we note that both codes can be defined over $\ff_{2^4}$, however by reducing the minimum distance from 
$7$ to $6$ we managed to reduce the locality by a factor of two, from $6$ to $3$. }
\end{xmpl}  

The additive and the multiplicative structures of the field can be combined into a more 
general method of constructing
good polynomials. 
For two subsets $H,G\subset \ff_q,$ we say that $H$ is closed under multiplication by $G$, if multiplying
elements of $H$ by elements of $G$ does not take the result outside $H$, i.e., if 
$\{hg:h\in H,g\in G\}\subseteq H$.
\begin{theorem} Let $l,s,m$ be integers such that $l$ divides $s$, $p^l\,\text{\rm mod\,} m=1$, and $p$ is a prime. Let $H$ be an additive subgroup of the field $\ff_{p^s}$ that is closed under the multiplication by the field $\ff_{p^l}$, and let $\alpha_1,...,\alpha_m$ be the $m$-th degree roots of unity in $\ff_{p^s}.$ Then for any $b\in \ff_{p^s}$ the polynomial 
\begin{equation}
g(x)=\prod_{i=1}^m\prod_{h\in H}(x+h+\alpha_i)
\label{eq:tjhk}
\end{equation}
is constant on the union of cosets of $H$,
$\cup_{1\le i\le m}H+b\alpha_i$, and the size of this union satisfies   
\begin{equation*}
|\cup_{1\le i\le m}H+b\alpha_i|=
\begin{cases}
|H| & \text{if }b\in H \\
m|H| & \text{if }b\notin H.
\end{cases}
\label{eq:imim}
\end{equation*}
\label{tytyw}
\end{theorem}

\begin{IEEEproof}
Let $\overline{h}\in H$ and let $\overline{h}+b\alpha_j$ be an arbitrary element, then
    \begin{align*}
g(\overline{h}+b\alpha_j)&=\prod_{i=1}^m\prod_{h\in H}(\overline{h}+b\alpha_j+h+\alpha_i)\nonumber\\
&=\prod_{i=1}^m\prod_{h\in H}(b\alpha_j+h+\alpha_i)\nonumber\\
&=\alpha_j^{-m|H|}\prod_{i=1}^m\prod_{h\in H}(b+h\alpha_j^{-1}+\alpha_i\alpha_j^{-1})\nonumber\\
&=\prod_{i=1}^m\prod_{h\in H}(b+h\alpha_j^{-1}+\alpha_i) \nonumber\\
&=\prod_{i=1}^m\prod_{h\in H}(b+h+\alpha_i)\\
&=g(b),\nonumber
\end{align*}
where we have made changes of the variables and 
used the assumption that $H$ is closed under multiplication by any $m$-th degree root of unity, since it is closed multiplication by $\ff_{p^l}$.
For the last part regarding the size of the union of the cosets, consider two distinct $m$-th roots of unity $\alpha_i, \alpha_j$, then
   $$
   H+b\alpha_i=H+b\alpha_j \;\Leftrightarrow\; b(\alpha_i-\alpha_j)\in H \;\Leftrightarrow\; b\in H,
   $$
where the last step follows since $\alpha_i-\alpha_j$ is a nonzero element of $\ff_{p^l}$ and $H$ closed under multiplication by 
the elements of $\ff_{p^l}$.
\end{IEEEproof}

{\bf Remarks:} 
\begin{enumerate}
\item In order to construct a good polynomial using Theorem \ref{tytyw}, one needs to find an additive subgroup $H$ of $\ff_{p^s}$ that is closed under multiplication by $\ff_{p^l}$.
Note that since $l$ divides $s$, the field $\ff_{p^s}$ can be viewed as a vector space of dimension $s/l$ over the field $\ff_{p^l}.$ Therefore any subspace $H$ of dimension $1\leq t\leq s/l$ is in fact an additive subgroup of the field 
	$\ff_{p^s}$ that is closed under multiplication by $\ff_{p^l}$, and is of size 
	$|H|=(p^{l})^{t}=p^{tl}.$ 
	\item 
Since the degree of the polynomial $g(x)$ in \eqref{eq:tjhk} is $m|H|$, it is clear that it takes distinct values on different sets of the form $U=\cup_i H+b\alpha_i$. In other words, $g(x)$ partitions $\ff_{p^s}$ into $(p^{s}-|H|)/m|H|$ sets of size $m|H|$ and one set of size $|H|$, according to the values taken on
the elements of the field. Hence, over the field of size $p^{s}$, one can construct an optimal LRC code of length $n\le p^s$ such that
$m|H|$ divides $n.$
\end{enumerate}

Assume that one wants to construct an LRC code over a field of a specific characteristic $p$, e.g., $p=2$, 
then Theorem \ref{tytyw} gives a flexible method of constructing good polynomials for a large set of parameters. More specifically,
let $m$ be an integer not divisible by $p$, 
and let $l$ be the smallest integer such that $p^l\text{mod\,}m=1$ (note that $l\le\phi(m)$, where $\phi(\cdot)$ is Euler's totient function).
Then is it possible to construct a good polynomial that is constant on sets of size $mp^{t}$ for any integer $t$ which is a multiple of $l$. 
\begin{xmpl}{\rm Suppose that $p=7$ and the code parameters are $(n=28,r=13).$
To construct an optimal LRC code we need to construct a polynomial $g(x)$ that is constant on two disjoint sets of size $r+1=14$ over some extension of $\ff_{7}$. Write $14=2\cdot 7$ then $m=2,l=1$, moreover, using Theorem \ref{tytyw} one can construct the desired good polynomial over the field $\ff_{7^2}$. More precisely, following Remark (2) above, the polynomial $g(x)$ partitions the field of size $49$ into $3$ sets of size $14$ and one set of size $7$. 
Hence in order to construct a code of length $n=28$ one can choose any two out of the three sets of size $14$. 
Note that the dimension of the code can take any value $k\le nr/(r+1)=26.$}
\end{xmpl}

Let us summarize the constructions of good polynomials depending on the value of the parameters. Suppose that we would like to construct
a good polynomial over a field extension of $\ff_p$ that is constant on disjoint subsets of points
of size $mp^t,$
where $m$ and $p$ are coprime, then
\begin{enumerate}
	\item If $t=0$, one can use multiplicative subgroups of some field extension $\ff_{p^l}$ that satisfies $p^l\,\text{mod}\, m=1$;
	\item If $t>0$ and $m=1,$ one can rely on additive subgroups;
	\item If $t,m>1$ and $t$ is a multiple of $l$, where $l$ is the smallest integer such that $p^l\,\text{mod}\, m=1$, 
the construction is accomplished by combining the additive and multiplicative structures of the field as in Theorem \ref{tytyw}.
\end{enumerate}

There is one case where we are not able to construct good polynomials. For example, using the technique
discussed above it is not possible to construct a code with locality $r=5$ over any extension of the field $\ff_2$. 
This follows since the size of the set is $r+1=5+1=3\cdot 2$, hence $m=3$ and $l=2$ is the smallest integer such that $2^l\,\text{mod}\, 3=1$, however $t=1$ is not a multiple of $l=2$. 
On the other hand, a simple counting argument shows
that good polynomials exist also for this unresolved case if the field $\ff_q$ is large enough.
\begin{proposition}
Let $\ff_q$ be the finite field of size $q$. There exists a good polynomial of degree $r+1$ that is constant on at least
  $\lceil\binom q{r+1}/{q^r}\rceil$ sets of size $r+1.$
  \end{proposition}
  \begin{IEEEproof}
Consider the set $M_{q,r}=\{f\in \ff_q[x]: f=\prod_{i=1}^{r+1}(x-\alpha_i)\},$ where $\alpha_i,i=1,\dots,r+1$ vary over
all $\binom q{r+1}$ possible choices of subsets of the field of size $r+1$. In other words, $M_{q,r}$ is the set of all monic polynomials of degree $r+1$ in $\ff_q[x]$ that also have $r+1$ \emph{distinct} zeros in $\ff_q$. We say that two polynomials $f(x)=x^{r+1}+\sum_{i=0}^r a_ix^i,g(x)\in M_{q,r},$ are equivalent if they differ by a constant. Clearly this is an equivalence relation on $M_{q,r}$, and the number of equivalence classes is at most $q^r$  according to the number of choices of $r$-tuples of the
coefficients $a_1,\dots,a_r.$ Hence there exists an equivalence class of size at least $\lceil\binom q{r+1}/{q^r}\rceil$.
Let $f$ be a representative of this class, and note that it is constant on the set of zeros of any other polynomial $g$ from this class. 
We conclude that $f$ is a good polynomial that is constant on sets of size $r+1$, and the number of sets is at least 
$\lceil \binom q{r+1}/q^r \rceil$.
\end{IEEEproof}


When $q$ is large enough, e.g., $q>n(r+1)^{r}$, the quantity $\lceil\binom q{r+1}/{q^r}\rceil$ exceeds 
$n/(r+1)$  which is the desired number of sets for the construction.
For instance, taking $q=2^{11},$ we observe that there exists a polynomial $g\in\ff_q[x]$ 
of degree $r+1=6$ that is constant on at least $3$ disjoint sets of size $6$. Indeed, we find that 
$\frac{\binom{2^{11}}{6}}{(2^{11})^5}\approx 2.82.$
Using Construction \ref{cnstr3} and the polynomial $g$, we can construct an optimal LRC code over $\ff_q$ 
of length $n=18$, locality $r=5$ and any dimension $k\leq 15$.

 \subsection{A general view of the LRC code family}
 In this section we study the mapping from the set of polynomials of the form \eqref{eq:97979} to $\ff^n$,
 generalizing the code construction presented above.
  
Let $A\subset\ff$, and let $\cA$ be a partition of $A$ into $m$ sets $A_i$. 
Consider the set of polynomials $\ff_{\cA}[x]$ of degree less than $|A|$ that are constant on the blocks of the partition: 
  \begin{align}\label{eq:algebra}
  \ff_{\cA}&[x]=\{f\in\ff[x]: \nonumber\\&f \text{ is constant on }A_i, i=1,\dots,m;\;\deg f< |A|\}.
   \end{align}

The {\em annihilator} of $A$ is the smallest-degree monic polynomial $h$ such that $h(a)=0$ if $a\in A,$ i.e., $h(x)=\prod_{a\in A} (x-a).$
Observe that the set $\ff_{\cA}[x]$ with the usual addition and multiplication modulo $h(x)$ becomes a commutative algebra
with identity. 
Since the polynomials $\ff_{\cA}[x]$ are constant on the sets of $\cA,$ 
we write $f(A_i)$ to refer to the value of the polynomial $f$ on the set $A_i\in \cA$.  
We will also use a short notation for multiplication of polynomials, writing $fg$ instead of $fg \,\text{mod\,}h.$  

The next proposition lists some properties of the algebra.
\begin{proposition}
\label{prop1}
\begin{enumerate}
\item Let $f\in\ff_{\cA}[x]$ be a nonconstant polynomial, then $\max_i|A_i|\le \deg (f)< |A|;$ 

\item The dimension $\dim(\ff_{\cA}[x])=m,$ and the $m$ polynomials $f_1,...,f_m$ that satisfy $f_i(A_j)=\delta_{i,j}$ and $\deg(f_i)<|A|,$
form a basis (here $\delta_{i,j}$ is the Kronecker delta). 
Explicitly, 
   \begin{equation}\label{timtam2}
f_{i}(x)=\sum_{a \in A_i}\prod_{b\in A\backslash a}\frac{x-b}{a-b}.
  \end{equation}

\item Let $\alpha_1,...,\alpha_m$ be distinct nonzero elements of $\ff$, and let $g$ be the polynomial of degree $\deg (g)<|A|$ that satisfies $g(A_i)=\alpha_i$ for all $i=1,...,m,$ i.e.,
       $$
 g(x)=\sum_{i=1}^m \alpha_i \sum_{a \in A_i}\prod_{b\in A\backslash a}\frac{x-b}{a-b} .
       $$
Then the polynomials $1,g,...,g^{m-1}$ form a basis of  $\ff_{\cA}[x].$
\item There exist $m$ integers  $0=d_0<d_1<...<d_{m-1}<|A|$ 
such that the degree of each polynomial in $\ff_{\cA}[x]$ is  $d_i$ for some $i$. 
\end{enumerate}
\end{proposition}

\begin{IEEEproof}

 (1) For a polynomial $f\in \ff_{\cA}[x]$, and a set $A_i\in\cA$, the polynomial $f(x)-f(A_i)$ has at least $|A_i|$ zeros in 
 $\ff,$ and therefore $\deg(f)\ge |A_i|.$ 
	
\vspace*{.05in} (2)  
The $m$ polynomials $f_1,...,f_m$ defined in \eqref{timtam2} are clearly linearly independent since if for some $\lambda_i$'s in the field, 
$$\sum_{i=1}^m\lambda_if_i(x)=0,$$
then for any $j=1,...,m$
   $$
   \sum_{i=1}^m\lambda_if_i(A_j)=\sum_{i=1}^m\lambda_i\delta_{i,j}=\lambda_j=0.
   $$
By definition, the polynomials $f_1,\dots,f_m$ span $\ff_{\cA}[x].$

\vspace*{.05in} (3) Because of part (2) it is sufficient to show that the polynomials $1,g,...,g^{m-1}$ are linearly independent.
Assume that for some $\beta_j$'s in $\ff$, 
\begin{equation}
\sum_{j=1}^{m}\beta_jg^{j-1}(x)=0.
\label{timtam3}
\end{equation}
Define the $m\times m$ matrix $V=(v_{i,j})$ where $v_{i,j}=(g^{j-1}(A_i)).$ From \eqref{timtam3} we conclude that $V\cdot(\beta_1,...,\beta_m)^T=0$, 
however $V$ is a Vandermonde matrix defined by $m$ distinct nonzero elements of the field, 
therefore it is invertible, and $\beta_i=0$ for all $i$.

(4) Let $f_0,...,f_{m-1}$ be a basis for the algebra $\ff_{\cA}[x]$. W.l.o.g. we can assume that the degrees of the polynomials are all distinct, since if this is not the case, one can easily find such basis by using linear operations on the $f_i$'s. 
For this, consider an $m\times |A|$ matrix whose rows are formed by the coefficient vectors of the polynomials $f_i$. 
The rows of the reduced row-echelon form of this matrix correspond to a basis of polynomials of distinct degrees.
Let $d_i=\deg(f_i)$, and assume that $d_0<d_1<...<d_{m-1}$. Since the constant polynomials are contained in the algebra, $d_0=0$, 
and the result follows. 
\end{IEEEproof}

Next we consider a special case of an algebra generated by a set $A$ of size $n,$ assuming that the
partition satisfies $|A_i|=r+1$ for all $i.$
\begin{corollary}
\label{cor1}
Assume that $d_1=r+1$, namely there exists a polynomial $g$ in $\ff_{\cA}[x]$ of degree $r+1$, 
then $d_i=i(r+1)$ for all $i=0,...,m-1$, 
 and the polynomials $1,g,...,g^{m-1}$ defined in Proposition \ref{prop1} part (3), form a basis for $\ff_{\cA}[x]$.  
\end{corollary}

\begin{IEEEproof}
If there exists such a polynomial $g$, then clearly it takes distinct values on distinct sets of the partition $\cA$. Otherwise for some constant $c\in \ff,$ the polynomial $g-c$ has at least $2(r+1)$ roots, and is of degree $r+1$, which is a contradiction. Hence, by Proposition \ref{prop1}, part (3) the powers of $g$ form a basis of the algebra, and the result follows. 
\end{IEEEproof}

Note that the algebra $\ff_{\cA}[x]$ in Construction \ref{cnstr3} contains a good polynomial of degree $r+1$, satisfying the
assumptions of Corollary \ref{cor1}, and therefore $\ff_{\cA}[x]$ is generated by the powers of this polynomial.

\vspace*{.1in}
Next let us use the properties of the algebra of polynomials defined by the partition $\cA$ 
to construct $(n,k,r)$ LRC codes.  
\begin{cnstr}
\label{cnstr1.1}{\rm
Let $A\subset\ff, |A|=n$ and let $\cA$ be a partition of the set $A$ into $m=\frac{n}{r+1}$ sets of size $r+1$.
Let $\Phi$ be an injective mapping from $\ff^k$ to the space of polynomials 
    $$
    \cF_{\cA}^r=\oplus_{i=0}^{r-1} \ff_{\cA}[x]x^i.
    $$
(Note that $\cF_{\cA}^r$ is indeed a direct sum of the spaces, so $\dim(\cF_{\cA}^r)=mr.$ Therefore such an injective mapping exists 
iff $k\leq mr=nr/(r+1)$).

The mapping $\Phi$ sends the set of messages $\ff^k$ to a set of encoding polynomials. 
We construct a code by evaluating the polynomials $f\in \Phi(\ff^k)$ at the points of $A$. If $\Phi$ is a linear mapping, then the resulting code is also linear.}
\end{cnstr}
This construction relies on an arbitrary mapping $\Phi:\ff^k\to\cF_{\cA}^r$. It forms a generalization of Construction \ref{cnstr3} which used a particular linear mapping for the same purpose.

Below we write $f_a(x):=\Phi(a).$

\begin{theorem}
\label{thm1} 
Construction \ref{cnstr1.1} gives an $(n,k,r)$ LRC code with minimum distance $d$ satisfying 
   \begin{equation}\label{eq:md}
   d\geq n-\max_{a,b\in \ff^k}\deg(f_a-f_b)\ge n-\max_{a\in \ff^k}\deg(f_a).
   \end{equation}
\remove{\textcolor{blue}{I have a problem with the claim about the distance. Since the mapping $\Phi$ may not be linear, therefore it might not be closed under addition. namely for two vectors 
$f_a,f_b\in \Phi(\mathbb{F}^k)$ not necessarily we have $f_a-f_b\in \Phi(\mathbb{F}^k)$. However since degree of $f_a-f_b$ is less than $n$ the distance is at least $1$. Therefore the correct bound should be 
$$d\geq n-\max_{a,b\in \ff^k}\deg(f_a-f_b)\}\geq n-\max_{a,b\in \ff^k}\deg(f_a-f_b)\}.$$
Am I right?}}
\end{theorem}

\begin{IEEEproof}
To prove local recoverability, we basically repeat the proof of Theorem \ref{thm1.1}. For a given message vector $a$ 
let 
\begin{equation}
f_a(x)=\sum_{i=0}^{r-1}f_i(x)x^i, 
\label{eq:97976}
\end{equation}
where the coefficient polynomials $f_i(x)$ satisfy $f_i\in\ff_{\cA}[x].$
Choose $j\in\{1.\dots,m\}$ and suppose that the symbol to be recovered is $f_a(\alpha),$ where
$\alpha\in A_j.$
 Define the decoding polynomial
\begin{equation}
\delta(x)=\sum_{i=0}^{r-1}f_i(\alpha)x^i
\label{eq:decoding}
\end{equation} 
and note that $\delta(\alpha)=f_a(\alpha)$ on account of \eqref{eq:97976}, \eqref{eq:decoding}.
Since $f_i$ belongs to $\ff_{\cA}[x]$, for any $\beta$ in $A_j$ we have $f_a(\beta)=\delta(\beta).$ 
Moreover, since $\delta(x)$ is of degree at most $r-1$, it can be interpolated by accessing the $r$ 
values of $f_a(\beta)=\delta(\beta)$ for $\beta$ in $A_j\backslash \alpha$. 
We conclude that the value of the lost symbol $f_a(\alpha)$ can be found by accessing the remaining $r$ symbols in the block $A_j.$

It remains to prove \eqref{eq:md}.  Let $(f_a(\alpha))_{\alpha \in A}$, $(f_b(\alpha))_{\alpha \in A}$ be two codewords constructed from distinct message vectors $a$ and $b$. Since $\Phi$ is injective and 
$\deg(f_a-f_b)<n,$ the code vectors that correspond to $f_a$ and $f_b$ are distinct. 
Then \eqref{eq:md} is immediate. \end{IEEEproof}

 \subsection{Systematic encoding of LRC codes}
 In implementations it is preferable to have a systematic form of LRC codes in order to easily retrieve the stored information.  
We note that all the constructions described above can be modified to yield systematic codes with no loss in the code distance,
by modifying the encoding polynomials \eqref{eq:97979}, \eqref{eq:97976}.
In particular Construction \ref{cnstr3} can be modified to give optimal LRC codes in a systematic form.
Such a modification is briefly described in this section.

Let $\cA=\{A_1,...,A_{m}\},\,m=n/(r+1)$ be a partition of the set $A\subseteq \ff$ of size $n$ into sets of size $r+1$. 
For $i=1,...,k/r$ let $B_i=\{\beta_{i,1},...,\beta_{i,r}\}$ be some subset of $A_i$ of size $r$. 
In our systematic encoding the message symbols will be written in the coordinates with locations in the sets $B_i.$

Recall that the algebra $\ff_\cA[x]$ has a basis of polynomials $f_i$ that satisfy $f_i(A_j)=\delta_{i,j}$ for $ i,j=1,...,m$
\eqref{timtam2}.
For each set $B_i$ define $r$ polynomials $\phi_{i,j},j=1,...,r$ of degree less than $r$ such that 
   $$
\phi_{i,j}(\beta_{i,l})=\delta_{j,l}.
   $$ 
These polynomials can be easily found using Lagrange's interpolation. 
For $k$ information symbols $a=(a_{i,j}), i=1,\dots,k/r; j=1,\dots,r$ 
define the encoding polynomial
\begin{equation}
f_a(x)=\sum_{i=1}^{k/r}f_i(x)\Big(\sum_{j=1}^ra_{i,j}\phi_{i,j}(x)\Big).
\label{54kl}
\end{equation}
The encoding of the message $a$ is defined by computing the vector ${(f_a(\alpha), \alpha\in A)},$ see \eqref{eq:code1}. 
It is easily verified that $f_a\in \cF_{\cA}^r,$ so each symbol has locality $r$. Furthermore, by definition
we have 
 $$
  f_a(\beta_{i,j})=a_{i,j}, \quad i=1,\dots,k/r; j=1,\dots,r,
  $$
so the code is indeed systematic.
  
Although \eqref{54kl} gives a systematic $(n,k,r)$ LRC code, optimality of the minimum distance is generally
not guaranteed. 
This follows since the best bound on the degree of the encoding polynomial $f_a(x)$ is $\deg(f_a)< n$. If the algebra $\mathbb{F}_\cA[x]$ is generated by the powers of a good polynomial $g$ (see Proposition \ref{prop1}, part (3)) then it is possible to construct an optimal systematic  LRC code. 
Indeed, one has to  
replace each polynomial $f_i$ in \eqref{54kl} with the polynomial $\overline{f_i}$ 
that is a linear combination of the polynomials $1,g,...,g^{(k/r)-1}$ and satisfies $\overline{f_i}(A_j)=\delta_{i,j}$ for all $j=1,\dots,k/r.$ 
This is possible since the matrix $V=(g^{j-1}(A_i))$ is a Vandermonde matrix and thus invertible.
Clearly the degree of each $\overline{f_i}$ is at most $((k/r)-1)(r+1)$. Therefore the degree of $f_a(x)$ is at most $k+(k/r)-2$, and  optimality of the distance follows.

\section{LRC codes with multiple recovering sets}\label{sect:multiple}
In this section we extend the original local recoverability problem in one more direction, requiring each symbol to
have more than one recovering set of $r$ symbols. 
Having in mind the applied nature of the problem, we will assume that the different recovering sets for the given symbol are disjoint. Indeed, in distributed storage applications there are subsets of the data that
are accessed more often than the remaining contents (they are termed ``hot data''). In the case that such segments
are accessed simultaneously by many users of the system, the disjointness property ensures that multiple read 
requests can be satisfied concurrently and with no delays. 

Let us give a formal definition. Let $\ff$ be a finite field. A code $\cC\subset \ff^n$ is said to be {\em locally recoverable with
$t$ recovering sets} (an LRC$(t)$ code) if for every $i\in\{1,\dots,n\}$ there exist disjoint subsets $A_{i,j}\subset [n]\backslash i, j=1,\dots,t$ of size $r_1,\dots,r_t$
respectively, such that for any codeword $x\in \cC,$ the value of the symbol $x_i$ is a function of each of the subsets 
of symbols $\{x_l, l\in A_{i,j}\}, j=1,\dots,t.$ We write $(n,k,\{r_1,\dots,r_t\})$ LRC code to refer to an LRC$(t)$ code of dimension $k$, 
length $n$, and $t$ disjoint recovering sets of size $r_i, i=1,\dots,t$.

We will present two methods of constructing LRC codes with multiple recovering sets, both relying on the construction of
the previous section. The first method relies on the combinatorial concept of orthogonal partitions, extending the basic
construction to multiple recovering sets. The second method uses the construction of product codes and graph codes to combine 
several LRC codes into a longer multiple recovering code. For simplicity of presentation we will restrict ourselves to codes with two recovering sets, although both constructions clearly apply for any number of recovering sets.

\subsection{Algebraic LRC codes with multiple recovering sets}
In this section we present a construction of LRC codes with multiple disjoint recovering sets 
that develops the method of Sect.~\ref{sect:construction}. 
As in the case for single recovering set, the construction will utilize the additive and multiplicative structure of the field. 

Let $A\subseteq\ff, |A|=n$ and let $\cA$ (respectively, $\cA'$) be a partition of $A$ into disjoint sets of size  $r+1$
(resp., $(s+1)$). Define two subspaces of polynomials 
   \begin{equation}\label{eq:Va}
\cF_{\cA}^r=\oplus_{i=0}^{r-1}\ff_{\cA}[x]x^i \;\;\text{ and }\;\; \cF_{\cA'}^s=\oplus_{i=0}^{s-1}\ff_{\cA'}[x]x^i,
   \end{equation}
where the notation $\ff_\cA[x]$ is defined in \eqref{eq:algebra}.
Clearly  
    $$
    \dim(\cF_{\cA}^r)=r\frac{n}{r+1},\quad \dim(\cF_{\cA'}^s)=s\frac{n}{s+1}.
    $$
For an integer $m$ let $\cP_m$ be the space of polynomials of degree less than $m$, and define 
  \begin{equation}\label{eq:Vm}
   V_m=\cF_{\cA}^r\cap \cF_{\cA'}^s\cap \cP_m
   \end{equation}
to be the space of polynomials of degree less than $m$ that also belong to $\cF_{\cA}^r\text{ and } \cF_{\cA'}^s$. 

\begin{cnstr}{\rm Let $A, \cA_1,\cA_2$ be as above. Assume that $\dim(\cF_{\cA}^r\cap \cF_{\cA'}^s)\ge k$ and
let $m$ be the smallest integer such that $\dim(V_m)=k$.
Let $\Phi:\ff^k \to V_m$ be an injective mapping. 
For simplicity we assume that this mapping is linear, i.e., 
there exists a polynomial basis $g_0,...,g_{k-1}$  of $V_m$ such that 
   $$
   \Phi(a)=\sum_{i=0}^{k-1}a_ig_i(x).
   $$
Denote by $f_a(x)=\Phi(a)$ the encoding polynomial for the vector $a$.
Construct the code as the image of $\ff^k$ under the evaluation map similarly to \eqref{eq:code1}.}
\label{multiple_recovering}
\end{cnstr}

Call partitions $\cA_1$ and $\cA_2$ {\em orthogonal} if 
    $$
      |X\cap Y|\le 1 \quad \text{for all } X\in \cA_1, Y\in \cA_2.
    $$
If the partitions $\cA_1$ and $\cA_2$ are orthogonal, then every symbol of the code constructed above has two \emph{disjoint} recovering sets of size $r$ and $s,$ respectively.

\begin{theorem}\label{thm:mr} Assume that the partitions in Construction \ref{multiple_recovering}  are orthogonal.
Then this construction gives an $(n,k,\{r,s\})$ LRC code $\cC$ with distance at least $n-m+1$.
\end{theorem}

\begin{IEEEproof} The claim about the distance is obvious from the construction (it applies even if the mapping $\Phi$ is nonlinear).
The local recoverability claim is proved as follows. 
Since the encoding polynomial $f_a$ is in $\cF_{\cA}^r$, there exist $r$ polynomials $f_0,...,f_{r-1}$ in $\ff_{\cA}[x]$ such that  
    $$
    f_a(x)=\sum_{i=0}^{r-1}f_i(x)x^i.
    $$
Now we can refer to  Theorem \ref{thm1}. Using the arguments in its proof, every symbol of the codeword can be
recovered by accessing the $r$ symbols from the block of the partition $\cA$ that contains it, as well as by accessing the $s$ symbols from
the corresponding block of the partition $\cA'.$
The result follows.
\end{IEEEproof}

In the following example we will construct an LRC$(2)$ code using Construction \ref{multiple_recovering} and two orthogonal partitions.
\begin{xmpl}{\rm
Let $\ff=\ff_{13},$ $A=\ff\backslash\{0\}$,  and let $\cA$ and $\cA'$ be the orthogonal partitions defined by the cosets of the multiplicative cyclic
groups generated by $5$ and $3$, respectively. 
We have 
\begin{equation}
\begin{array}{c}
\cA=\{\{1,5,12,8\}, \{2,10,11,3\}, \{4,7,9,6\} \}\\[.05in] \cA'=\{\{1,3,9\}, \{2,6,5\},\{4,12,10\}, \{7,8,11\}\}.
 \end{array}
\label{tre}
\end{equation}
Since $|\cA|=3$, by Proposition \ref{prop1}, $\dim(\ff_{\cA}[x])=3,$ and similarly, $\dim(\ff_{\cA'}[x])=4$. 
It is easy to check that 
    $$
    \ff_{\cA}[x]=\langle 1,x^4,x^8\rangle,\quad \ff_{\cA'}[x]=\langle 1,x^3,x^6,x^9\rangle.$$
Moreover by \eqref{eq:Va}
\begin{align}
\cF_{\cA}^r\cap \cF_{\cA'}^s&=\langle 1,x,x^2,x^4,x^5,x^6,x^8,x^9,x^{10}\rangle\nonumber
\\&\hspace*{.5in}\cap \langle 1,x,x^3,x^4,x^6,x^7,x^9,x^{10}\rangle\nonumber \\
&=\langle 1,x,x^4,x^6,x^9,x^{10}\rangle. \label{gfgfy}
\end{align}
Let $m=7,$ then 
\begin{equation}
V_m=\langle 1,x,x^4,x^6\rangle.
\label{eq:wmwm}
\end{equation}

We will construct a $(12,4,\{2,3\})$ LRC code with distance $d\ge 6$. 
By Construction \ref{multiple_recovering} and \eqref{eq:wmwm}, 
for a vector $a=(a_0,a_1,a_2,a_3)\in\ff^4$ the encoding polynomial is 
     $$
     f_a(x)=a_0+a_1x+a_2x^4+a_3x^6.
     $$
This polynomial can be written as 
   \begin{gather*}
   f_a(x)=\sum_{i=0}^{2}f_i(x)x^i, 
   \\
   \text{ where } f_0(x)=a_0+a_2x^4, \;f_1(x)=a_1, \;f_2(x)=a_3x^4,
   \end{gather*}
and each $f_i\in \ff_{\cA}[x]$.
The same polynomial can also be written as 
    \begin{gather*}
f_a(x)=\sum_{i=0}^{1}g_i(x)x^i\\
\text{ where } g_0(x)=a_0+a_3x^6, g_1(x)=a_1+a_2x^3,
    \end{gather*}
and $g_0,g_1\in \ff_{\cA'}[x]$.

Assume that one would like to recover the value of the codeword symbol $f_a(1).$ This can be done in
two ways as follows:

(1) Use the set  in the partition $\cA$ that contains $1$, i.e., $\{1,5,12,8\},$ find the polynomial 
$\delta(x)$ of degree at most $2$ such that $\delta(5)=f_a(5),\delta(12)=f_a(12)$ and $\delta(8)=f_a(8).$
The symbol $f_a(1)$ is found as $f_a(1)=\delta(1);$

or 

(2) Use the set $\{1,3,9\}\in \cA',$ which also contains 1, find the polynomial $\delta_1(x)$ 
of degree at most $1$ such that $\delta_1(3)=f_a(3),\delta_1(9)=f_a(9)$.
The symbol $f_a(1)$ is found as $f_a(1)=\delta_1(1)$.  

Finally, since $\deg f_a\le 6$ for all $a\in \ff_k,$ we immediately observe that $d(\cC)\ge 6.$}
\end{xmpl}

As observed above, orthogonality of the partitions is a desirable property in the context of simultaneous data recovery
by different users. In \eqref{tre} we constructed orthogonal partitions using cosets of two distinct subgroups of the
field $\ff$. Of course, not every pair of subgroups has this property.
It is easy to identify a necessary and sufficient condition for the subgroups to generate
orthogonal partitions. 
\begin{proposition}
Let $H$ and $G$ be two subgroups of some group, then the coset partitions $\cH$ and $\cG$ defined by $H$ and $G$ respectively are orthogonal iff the subgroups intersect trivially, namely 
$$H\cap G=1.$$ If the group $X$ is cyclic, then it is equivalent to requiring that $\gcd (|H|,|G|)=1.$
\label{sdf}
\end{proposition}

\begin{IEEEproof}
Two distinct elements $x,y$ in the group are in the same cosets in the partitions $\cH$ and $\cG$ iff $Hx=Hy$ and $Gx=Gy$, 
which is equivalent to $xy^{-1}\in H\cap G$ and $xy^{-1}\neq 1$, and the first part follows. Now assume that the group is 
cyclic (e.g. the multiplicative group of a finite field), and let $h=|H|$ and $g=|G|.$ 
Elements $x, y$ 
belong to the same coset in the partitions $\cH$ and $\cG$ 
iff the element $xy^{-1}$ is both an $h$-th and $g$-th root of unity. 
This happens if and only if the order $\ord(xy^{-1})$ divides both $h$ and $g$, or equivalently that $\ord(xy^{-1})|\gcd(h,g)$.
Since $x\ne y,$ the order $\ord(xy^{-1})>1$, hence $\gcd(h,g)\neq 1$, which proves the second part.
\end{IEEEproof}

In the context of finite fields we can use both the multiplicative group (as in the above example) and the additive group
of the field to construct LRC$(t)$ code.
\begin{xmpl} {\rm In applications it is often useful to have codes over a field of characteristic $2$, e.g., over the field $\ff_{16}.$
We have
$\ff_{16}^+\cong \ff_4^+\times \ff_4^+,$
and the two copies of $\ff_4^+$ in $\ff_{16}$ intersect
only by the zero element, hence by Proposition \ref{sdf} they generate two orthogonal partitions.
Using Construction \ref{multiple_recovering}, one can construct an LRC code of length $16$ with two 
disjoint recovering sets for each symbol, each of size $3$. The dimension of the code can be any integer $k\le 8.$}
\end{xmpl}

Since the additive group of the field is a direct product of smaller groups, it is easy to find subgroups that intersect trivially, 
giving rise to orthogonal partitions of $\ff_q.$ These partitions can be used to construct LRC$(2)$ codes 
with disjoint recovering sets, as in the previous example. 

At the same time, constructing LRC$(2)$ codes from a multiplicative subgroup
of $\ff_q, q=p^l$ requires one extra condition, namely, that $q-1$ is not a power of a prime. In this case,
we can find two subgroups of $\ff_q^\ast$ of coprime orders, which give rise to 
orthogonal partitions of  $\ff_q^\ast.$
\begin{proposition} Let $\ff_q$ be a finite field such that the $q-1$ is not a power of a prime.
Let $r,s>1, gcd(r,s)=1$ be two factors of $q-1.$ Then there exists an LRC$(2)$ code $\cC$ of length $q-1$ 
over $\ff_q$ such that every code symbol has two disjoint recovering sets of sizes $r-1$ and $s-1$. The code $\cC$ can be constructed using
Construction \ref{multiple_recovering}
based on the subgroups of $\ff_q^\ast$ of orders $r$ and $s.$
\end{proposition}
One sufficient condition for the existence of subgroups of coprime orders in the multiplicative group of $\ff_{p^l}$ is that $l$
itself is not a power of a prime. Indeed, let $l=ab,$ where $a\leq b$ and $a$ does not divide $b$. 
In this case both $(p^a-1)|(p^l-1)$
and $(p^b-1)|(p^l-1)$. 
Then $p^l-1$ is not a power of a prime, because otherwise $(p^a-1)|(p^b-1)$, i.e., $a|b.$

\begin{xmpl}{\rm 
Using Construction \ref{multiple_recovering} and the previous observation, one can construct an LRC$(2)$ code of length $2^6-1=63,$ 
in which every symbol has two disjoint recovering sets of size $2$ and $6$, respectively. 
This is done using the orthogonal partitions derived from the subgroups of size $3$ and $7$.}
\end{xmpl}

LRC codes with multiple disjoint recovering sets are likely to have large minimum distance since each 
erased symbol can be recovered in several ways, so the code is resilient against many erasures. 
In the following statement we quantify this argument by establishing a lower bound on the distance in terms of the number
of recovering sets for each symbol. The next theorem applies to any class of LRC$(t)$ codes such that the recovering
sets for the symbols form $t$ mutually orthogonal partitions.
\begin{theorem}\label{thm:dist}
Let $\cC$ be an LRC$(t)$ code of length $n$, and suppose that the recovering sets are given by mutually orthogonal partitions
$\cA_1,...,\cA_t$ of $[n].$ Let 
$m$ be the smallest positive integer that satisfies
    \begin{equation}\label{eq:987}
t f(m)\leq \binom{m}{2},
    \end{equation}
where
    $$
     f(m)=\begin{cases}
   \frac{m}{2}, & m \text{ even}\\
    \frac{m+3}{2}, & m \text{ odd.}
     \end{cases}
   $$ 
Then the distance of $\cC$ is at least $m$.
\end{theorem}
The proof relies on the following lemma.
\begin{lemma} \label{lemma:<}
Let $\cA_1,...,\cA_t$ be $t$ mutually orthogonal partitions of a finite set $A$, and let $m$ be defined in \eqref{eq:987}. 
Then for any $B\subset A, |B|< m$ there exists a subset $C$ in some partition $\cA_i, i=1,\dots,t$ such that 
$$|B\cap C| =1.$$
\label{thm:partition}
\end{lemma}
\begin{IEEEproof}
By definition of $m$, for any integer $s<m$  
\begin{equation}
t f(s)>\binom{s}{2}.
\label{tyty}
\end{equation}
Assume toward a contradiction that the statement is false, then for every $i=1,\dots,t$ and any element $x\in B$, there exists $y\in B$ 
such that $x,y$ belong to the same set in the partition $\cA_i$.
For a partition $\cA_i$ define the graph $G_i$ with the elements of $B$ as its vertices, and draw an edge between $x$ and $y$ iff they are in the same set in the partition $\cA_i$. By the assumption, the degree of every vertex of $G_i$ is at least one. 
If $s=|B|$ is even then there are at least $s/2$ edges in $G_i.$ If $s$ is odd, then $G_i$ contains at least one
triangle, and so there are at least $(s-3)/2+3=(s+3)/2$ edges in it. 
Notice that since the 
partitions are mutually orthogonal, there are no edges that are contained in more than one graph $G_i.$ Therefore
    $$
    t f(s)\leq \sum_{i=1}^t|E(G_i)|=|\cup_{i=1}^tE(G_i)|\leq \binom{s}{2},
    $$
which is a contradiction to \eqref{tyty}. 
\end{IEEEproof}

{\em Proof of Theorem \ref{thm:dist}:} 
In order to prove that $d(\cC)\ge m$ we will show that any $m-1$ erased symbols in the codeword can be recovered. 
Let $B$ be the set of $m-1$ erased coordinates. By Lemma \ref{thm:partition} there exists a set $C$ in some partition $\cA_i$ 
such that $B\cap C=\{i_1\},$ where $i_1\in[n]$ is some coordinate. Since no other coordinates in the set $C$ are erased,
this permits us to recover the value of the symbol in the coordinate $i_1$ by accessing the symbols in  
the set $C\backslash \{i_1\}.$ This reduces the count of erasures by 1, leaving us with the set of erasures of cardinality $m-2$.
Lemma \ref{lemma:<} applies to it, enabling us to correct one more erasure, and so on.  \IEEEQED

\vspace*{.1in}
Let us show that Theorem \ref{thm:dist} can sometimes provide a better bound on the minimum distance compared
to the degree estimate.

\begin{xmpl}{\rm
Consider an $(n=12,k=6,\{r_1=2,r_2=3\})$ LRC code $\cC$ over $\ff_{13}$ obtained using Construction \ref{multiple_recovering}, the partitions in \eqref{tre}, and the corresponding algebras $\ff_{\cA}[x],\ff_{\cA'}[x]$. 
Using \eqref{eq:987} in Theorem \ref{thm:dist} we find that the distance of $\cC$ is at least $4$.

By \eqref{gfgfy} the set $\{1,x,x^4,x^6,x^9,x^{10}\}$ forms a basis of the space of encoding polynomials. 
Given a message vector $a=(a_0, a_1,a_4,a_6,a_9,a_{10})\in\ff^6$, write the encoding polynomial as
   $$
   f_{a}(x)=a_0+a_1x+a_4x^4+a_6x^6+a_9x^9+a_{10}x^{10}.
   $$
To find the codeword, evaluate the polynomial at all nonzero elements of the field $\ff_{13}$. 

Assume that the value $f_a(2)$ is erased and needs to be recovered. This can be done in two ways:

(1) Write the encoding polynomial as follows
    \begin{multline*}
f_a(x)=(a_0+a_4x^4)+x(a_1+a_9x^8)+ x^2(a_6x^4+a_{10}x^8)\\=g_0+g_1(x)x+ g_2(x)x^2,
    \end{multline*}
where $g_0=a_0+a_4x^4$, $g_1(x)=a_1+a_9x^8$, $g_2(x)=a_6x^4+a_{10}x^8,$ and $g_i\in\ff_{\cA}[x],i=1,2,3.$  
The symbol $f_a(2)$ can be found from the values of $f_a(10),f_a(11),f_a(3).$

(2) Write the encoding polynomial as follows
    \begin{multline*}
f_a(x)=(a_0+a_6x^6+a_9x^9)+x(a_1+a_4x^3+a_{10}x^9)\\=f_0(x)+xf_1(x),
     \end{multline*}
where $f_0(x)=a_0+a_6x^6+a_9x^9$ and $f_1(x)=a_1+a_4x^3+a_{10}x^9$, and $f_0,f_1\in\ff_{\cA'}[x]$. 
The symbol $f_a(2)$ can be found from the values of $f_a(5),f_a(6).$ 

Since the polynomial $f_a$ in this example can be of degree 10,
bounding the codeword weight by the degree  would only give the estimate $d(\cC)\ge 2.$}
\end{xmpl}

{\em Remark:}
As discussed above, an obvious solution to the multi-recovery problem is given by repeating each symbol of the data several
times. An advantage of this is high availability of data: Namely, a read request of a data fragment 
located on an unavailable or overloaded (hot) node can be easily satisfied by accessing the other replicas of the data. 
The LRC$(2)$ code $\cC$ constructed in the above example can be a good candidate to replace the repetition code, with almost no extra cost. 
Indeed, both the $(12,6)$ LRC$(2)$ code $\cC$ and the $(18,6)$ three-fold repetition code encode $6$ information symbols, 
however the encoding $\cC$ entails a $100\%$ overhead compared to a $200\%$ overhead in the case of repetition. 
The code $\cC$ is resilient to any $3$ erasures while the repetition code can fail to recover the 
data if all the 3 copies of the same fragment are lost. 
At the same time, the code $\cC$ uses subsets of sizes $2$ and $3$ to calculate the value of the symbol while
the repetition code in the same situation uses two subsets of size 1. Thus, the reduction of the overhead is attained at the
expense of a small amount of added computation.

In the final part of this section we derive a bound on the distance of the constructed codes confining ourselves to the basic 
case of the $(n,k,\{r,r\})$ code. This is accomplished by estimating the dimension of the subspace $V_m$ defined in \eqref{eq:Vm}
and then using Theorem \ref{thm:mr}.
\begin{lemma} Let $A$ be a set of size $n$, and assume that $\cA$ and $\cA'$ are two orthogonal partitions of $A$ into 
subsets of size $r+1$. Suppose that there exist polynomials $g$ and $g'$ of degree $r+1$ 
that are constant on the blocks of $\cA$ and $\cA',$ respectively. Then the dimension of the space $V_m$ \eqref{eq:Vm}
is at least $m(r-1)/(r+1).$     
\end{lemma}
\begin{IEEEproof}
Recall the space of polynomials $\cF_{\cA}^r$ defined in \eqref{eq:Va}.
Let $t=n/(r+1)$ and note that the basis of this subspace is given by the polynomials $g^ix^j,i=0,...,t-1 , j=0,...,r-1.$ 
Next we argue that
   $$
    \cP_m=(\cF_{\cA}^r\cap \cP_m)\oplus\text{Span}\,\Big\{x^{j(r+1)-1}, j=1,\dots,\Big\lfloor\frac m{r+1}\Big\rfloor\Big\},
  $$
  so 
  $$
  m=\dim(\cF_{\cA}^r \cap \cP_m)+\Big\lfloor\frac m{r+1}\Big\rfloor.
  $$
Thus, for any integer $m$,
    $$
\dim(\cF_{\cA}^r \cap \cP_m)\geq  \frac{mr}{r+1}
    $$ 
and the same bound holds if $\cA$ on the previous line is replaced with $\cA'.$
Then we obtain
\begin{align*}
\dim(\cP_m)&=m\\
&\geq \dim((\cF_{\cA}^r \cap \cP_m)+(\cF_{\cA'}^r \cap \cP_m))\\
&\geq \frac{mr}{r+1}+\frac{mr}{r+1}-\dim(\cF_{\cA}^r \cap \cF_{\cA'}^r\cap \cP_m)
\end{align*}
(cf. \eqref{eq:Vm}). Solving for the dimension of the subspace $\cF_{\cA}^r \cap \cF_{\cA'}^r\cap \cP_m=V_m$, we obtain the claimed estimate.
\end{IEEEproof}

Now suppose we have an $(n,k,\{r,r\})$ LRC code designed using Construction \ref{multiple_recovering}.
Choosing $m=\lceil \frac{k(r+1)}{r-1}\rceil$ we observe that the dimension of $V_m$ is at least $k.$ 
Therefore, from Theorem \ref{thm:mr} the distance of the code satisfies the inequality
    \begin{equation}\label{eq:d2}
    d\geq n-\Big(\Big\lceil \frac{k(r+1)}{r-1}\Big\rceil-1\Big)=n-k-\Big\lceil \frac{2k}{r-1}\Big\rceil+1.
    \end{equation}

{\em Remark:} While the paper was in review, a new bound on codes with multiple recovering sets was proved in \cite{mr}.
Using this result, we obtain the following inequalities for the distance $d$ of 
an $(n,k,\{r,r\})$ code:
   $$
    n-k-\Big\lceil \frac{2k}{r-1}\Big\rceil+1 \le d \le n-k-\Big\lfloor\frac{k-1}{r}\Big\rfloor-\Big\lfloor\frac{k-1}{r^2}\Big\rfloor+1.
   $$

\subsection{LRC Product Codes}
Given a set of $t$ LRC codes, one can construct an LRC$(t)$ code by taking a product of the corresponding linear subspaces.
Again for simplicity we confine ourselves to the case of $t=2$.
\begin{cnstr}
\label{cnstr5} {\rm We construct an $(n,k,\{r_1,r_2\})$ LRC code with $n=n_1n_2,k=k_1k_2$ by combining two LRC codes 
with the parameters $(n_i,k_i,r_i), i=1,2$  obtained by Construction \ref{cnstr1.1}. Suppose that the codes 
$\cC_1$ and $\cC_2$ are linear, and were constructed using linear injective mappings $\Phi_i$ and evaluating sets $A_i\in \ff, i=1,2$. 
Define the linear mapping 
$$\Phi=\Phi_1 \otimes \Phi_2: \ff^{k_1k_2}\rightarrow \oplus_{i=0}^{r_1-1}\ff_{\cA_1}[x]x^i\otimes \oplus_{j=0}^{r_2-1}\ff_{\cA_2}[y]y^j,$$
which is the tensor product of the mappings $\Phi_i$. 
Define the encoding polynomial for $a\in \ff^{k_1k_2}$ to be 
$$f_a(x,y)=\Phi(a).$$
The code is the image of $\ff^k$ under the evaluation map applied on the set of pairs  $A_1\times A_2$.}
\end{cnstr}
The following simple proposition summarizes the properties of this construction.
\begin{proposition}
Let $\cC_i\subset \ff^{n_i}$ be an $(n_i,k_i,r_i)$ LRC code with minimum distance $d_i, i=1,2.$ Construction \ref{cnstr5}
yields an LRC$(2)$ code with the parameters $(n=n_1n_2,k=k_1k_2,\{r_1,r_2\})$ and distance $d= d_1d_2$ .
\end{proposition}
\begin{IEEEproof} Denote by $\cA_i=\sqcup_{j\ge 1}A_j^{(i)}$ the partitions of the evaluation sets used in constructing the codes $\cC_i,i=1,2$ (refer
to Construction \ref{cnstr1.1}).
Let $a\in \ff^k$ and let the corresponding encoding polynomial be $f_a(x,y).$
Suppose that, for some point  $(x_0,y_0)\in A_1\times A_2$ we would like to compute in two ways the value of $f_a(x_0,y_0)$
by accessing $r_1$ and $r_2$ symbols, respectively.
Observe that the univariate polynomial $f_a(x,y_0)$ is contained in $\oplus_{i=0}^{r_1-1}\ff_{\cA_1}[x]x^i$, and therefore $f_a(x_0,y_0)$
can be found from the symbols in the set $\{f_a(\alpha,y_0), \alpha \in A_{m}^{(1)}\backslash x_0\},$ where 
$A_m^{(1)}\in \cA_1$ is the set that contains $x_0$.
Similarly $f_a(x_0,y_0)$ can be recovered using the polynomial $f_a(x_0,y)$ and the symbols in the set 
$\{f_a(x_0,\beta),\beta \in A_l^{(2)}\backslash y_0\},$ where $A_l^{(2)}\in\cA_2$ is the set that contains $y_0.$ Hence, 
the symbol $f_a(x_0,y_0)$ has two disjoint recovering sets of size $r_1,r_2,$ and the result follows.
\end{IEEEproof}

For instance, taking two optimal component LRC codes $\cC_1$ and $\cC_2$ with the parameters $(n_i,k_i,r), i=1,2$ we find the distance
of their product to satisfy
    \begin{equation}\label{eq:dp}
d= \Big(n_1-k_1-\Big\lceil \frac{k_1}{r}\Big\rceil+2\Big)\Big(n_2-k_2-\Big\lceil \frac{k_2}{r}\Big\rceil+2\Big)
     \end{equation}

\begin{xmpl} Let us construct an $(81,16,\{2,2\})$ LRC code $\cC\otimes\cC,$ where  $\cC$
is the optimal $(9,4,2)$ LRC code constructed in Example \ref{xmpl1}. 
The encoding polynomial of $\cC$ for a vector $a\in (\ff_{13})^4$ is $$f_a(x)=a_0+a_1x+a_2x^3+a_3x^4.$$
Define the vector $(b_0,b_1,b_2,b_3)=(0,1,3,4)$ and note that $f_a$ can be written as $f_a(x)=\sum_{i=0}^{3}a_ix^{b_i}.$
For a vector $a\in (\ff_{13})^{16}, a=(a_{i,j}),i,j=0,...,3$ the encoding polynomial of the product code $\cC \otimes \cC$ is 
$$f_a(x,y)=\sum_{i,j=0}^3a_{i,j}x^{b_i}y^{b_j}.$$ 
The codeword that corresponds to the message $a$ is obtained by evaluating $f_a$ at the points of $A\times A,$ where 
$A=\{1, 3, 9, 2, 6, 5, 4, 12, 10\}.$ 

Assume that the symbol $f_a(1,2)$ is erased and needs to be recovered. We can do it in two ways:

(1) Find the polynomial $\delta(x), \deg \delta(x)\le 1$ such that $\delta(3)=f_a(3,2),\delta(9)=f_a(9,2),$ and compute $f_a(1,2)=\delta(1)$,
or

(2) Find the polynomial $\delta_1(y), \deg \delta_1(y)\le 1$ such that $\delta_1(6)=f_a(1,6),\delta_1(5)=f_a(1,5),$ and compute $f(1,2)=\delta_1(2)$.
\end{xmpl}

We remark that product codes can be also viewed as codes on complete bipartite graphs. Replacing the complete
graph with a general bi-regular graph, we obtain general bipartite graph codes. A bipartite graph code is a linear code in which
the coordinates of the codeword are labeled by the edges of the graph, and a vector is a codeword if and only if the edges
incident to every vertex satisfy a given set of linear constraints. For instance, if this set is the same for every vertex
(and the graph is regular), we obtain a graph code in which the local constraints are given by some fixed code $\cC_0$
of length equal to the degree $\Delta$ of the graph. Having in mind our goal of constructing LRC codes, we should take $\cC_0$ 
to be a single erasure-correcting code of length $\Delta.$ This will give us a code with two recovering sets for every symbol, given by the vertices
at both ends of the corresponding edge. The advantage of this construction over product codes is that the length $\Delta$ of the
component code can be small compared to the overall code length $n$. We will confine ourselves to these brief remarks, 
referring the reader to the literature (e.g., \cite{bz05}) for more details on bipartite graph codes including estimates
of their parameters.

\vspace*{.1in}
{\em Comparing the two methods:}
The most fundamental parameter of an erasure-correcting code is the minimum distance.
To compare the two constructions, suppose that the desired parameters of the LRC$(2)$ code are $(n,k,\{r,r\})$ LRC codes
and use the expressions \eqref{eq:d2} and \eqref{eq:dp}. For simplicity, let us compare the constructions in terms of the
rate $R=k/n$ and the normalized distance $\theta=d/n.$ Then for Construction \ref{multiple_recovering} we obtain 
   $$
   \theta\ge 1-R\frac{r+1}{r-1} +O(1/n)
   $$
while for the product construction (Construction \ref{cnstr5}) we obtain \eqref{eq:dp}
  $$
 \theta = \Big(1-R_1\frac{r+1}{r}+O(1/n)\Big)\Big(1-R_2\frac{r+1}{r}+O(1/n)\Big)
  $$
Putting $R_1=R_2=\sqrt R$ gives the largest value on the right, and we obtain
   $$
  \theta= \Big( 1-\sqrt{R}\frac{r+1}{r}+O(1/n)\Big)^2
  $$
We observe that Construction \ref{multiple_recovering} gives codes with higher minimum distance 
than the
product of two optimal codes if the target code rate satisfies
  \begin{align*}
  R&\le \Big(\frac{2r(r-1)}{2r^2-1}\Big)^2=\Big(1-\frac 1r\Big)^2\Big(1+\frac 1{2r^2}+O\Big(\frac1{r^{4}}\Big)\Big)^2\\
  &\approx \Big(1-\frac 1r\Big)^2
  \end{align*}
(e.g., for $r=4$ the condition becomes $R\le 0.599$).

At the same time, the product construction provides more flexibility in constructing LRC codes with multiple recovering sets
because it gives multiple disjoint recovering sets by design.
On the other hand, Construction \ref{multiple_recovering} requires constructing several mutually orthogonal partitions with
their corresponding good polynomials, which in many cases can be difficult to accomplish.
Moreover, the product construction requires the field of size about $\sqrt n$, outperforming Construction \ref{multiple_recovering} 
which relies on the field of size about $n$, where $n$ is the code length.
Concluding, each of the two constructions proposed has its advantages and disadvantages, and therefore is
likely to be more suitable than the other one in certain applications.

\section{Generalizations of the Main Construction}\label{sect:generalizations}
In this section we return to the problem of LRC codes with a single recovering set for each symbol, generalizing
the constructions of Section \ref{sect:construction} in several different ways. We begin with constructing an LRC code 
for arbitrary code length, removing the assumption that $n$ is a multiple of $r+1$. 
We continue with a general method of constructing LRC codes with recovering sets of arbitrary given size, further
extending the results of Section \ref{sect:construction}.
One more extension that we consider deals with constructing optimal LRC codes in which each symbol is contained in a local code 
with large minimum distance.

\subsection{Arbitrary code length}\label{sect:arbitrary}
The constructions of Section \ref{sect:construction} require the assumption that $n$ is a multiple of $r+1.$ 
To make the construction more flexible, let us modify the definition of the codes so that this constraint is relaxed.
While the minimum distance of the codes presented below does not always meet the Singleton-type 
bound \eqref{eq:erer}, we will show that 
for the case of linear codes it is at most one less than the maximum possible value. 
The only assumption that will be needed is that $n\mod (r+1)\neq 1.$ 

As before, for $M\subset\ff$ denote by 
$h_{M}(x)=\prod\limits_{\alpha \in M}(x-\alpha)$
the annihilator polynomial of the set $M$. 
In the following construction we assume that $n$ is not a multiple of $r+1$.
For simplicity we also assume that $r$ divides $k$ although this constraint can be easily lifted at the expense of a 
somewhat more complicated notation.
\begin{cnstr} \label{cnstr_n_r}
{\rm Let $\ff$ be a finite field, and let $A\subset\ff$ be a subset such that $|A|=n, n\,\text{mod} (r+1)=s\neq 1$.
Let $m=\lceil\frac{n}{r+1}\rceil$ and let $\cA=\{A_1,\dots,A_m\}$ be a partition of $A$ 
such that $|A_i|=r+1,1\le i\le m-1$ and $1<|A_m|=s<r+1.$ 
Let $\Phi_i: \ff^{k/r} \to \ff_{\cA}[x], i=0,\dots,r-1$ be injective mappings. 
Moreover, assume that $\Phi_{s-1}$ is a mapping to the subspace of polynomials of $\ff_{\cA}[x]$ 
that vanishes on the set $A_m$, i.e., 
the range of $\Phi_{s-1}$ {is the space $\{f\in \ff_{\cA}[x]:f(\alpha)=0\text{ for any } \alpha\in A_m\}$.}
 
Given the input information vector $a=(a_0,...,a_{r-1})\in\ff^k$, where each $a_i$ is a vector of dimension $k/r$, 
define the encoding polynomial as follows:
\begin{align}
f_a(x)&=\sum_{i=0}^{s-1}\Phi_i(a_i)x^i+\sum_{i=s}^{r-1}\Phi_i(a_i)x^{i-s}h_{A_m}(x)\nonumber\\
&=\sum_{i=0}^{s-1}f_i(x)x^i+\sum_{i=s}^{r-1}f_i(x)x^{i-s}h_{A_m}(x),
\label{eq:97978}
\end{align}
where $\Phi_i(a_i)=f_i(x)\in\ff_{\cA}[x]$. Finally, define the code as the image of the evaluation mapping
similarly to \eqref{eq:code1}.}
\end{cnstr}

\begin{theorem}
Construction \ref{cnstr_n_r} defines an $(n,k,r)$ LRC code.
\end{theorem}

\begin{IEEEproof} Any symbol $f_a(\alpha)$ for $\alpha$ in one of the sets $A_1$,...,$A_{m-1}$ can be locally recovered using
the same decoding procedure as in Construction \ref{cnstr1.1}. 
This follows since the encoding polynomial $f_a(x)$ belongs to the space $\oplus_{i=0}^{r-1} \ff_{\cA}[x]x^i$, and therefore this 
symbol can be recovered by accessing $r$ symbols.
The only special case is recovering symbols in the set $A_m.$ By definition of $\Phi_{s-1}$ and \eqref{eq:97978}, the restriction of the encoding polynomial
 $f_a(x)$ to the set $A_m$ is a polynomial of degree at most $s-2$. Hence in order to recover the value of $f_a(\alpha)$ 
 for an element $\alpha\in A_m$, we find the polynomial $\delta(x)=\sum_{i=0}^{s-2}f_i(\alpha) x^i$ from the set of $s-1$ values
$\delta(\beta)=f_a(\beta), \beta\in A_m\backslash \{\alpha\}.$ 
Clearly the lost symbol is $f_a(\alpha)=\delta(\alpha)$, and the locality property follows.
\end{IEEEproof}

To estimate the value of the code distance consider the following modification of Construction \ref{cnstr_n_r}.

\begin{cnstr}
\label{cnstr_n_r_linear}
{\rm
Let $\ff$ be a finite field, and let $A\subset\ff$ be a subset such that $|A|=n, n\,\text{mod} (r+1)=s\neq 0,1$.
Assume also that $k+1$ is divisible by $r$ (this assumption is nonessential).

Let $\cA$ be a partition of $A$ into $m$ subsets $A_1,\dots,A_m$ of sizes as in Construction \ref{cnstr_n_r}. Let $g(x)$ be a polynomial of degree $r+1$, 
such that its powers $1,g,...,g^{m-1}$ span the algebra $\ff_{\cA}[x].$ W.l.o.g. we can assume that $g$ vanishes on the set 
$A_m$, otherwise one can take the powers of the polynomial $g(x)-g(A_m)$ as the basis for the algebra. 

Let $a=(a_0,...,a_{r-1})\in \ff^k$ be the input information vector, such that each $a_i$ for 
$i\neq s-1$ is a vector of length $(k+1)/r$ and $a_{s-1}$ is of length $\frac{k+1}{r}-1.$ 
Define the encoding polynomial 
\begin{multline}
f_a(x)=\sum_{i=0}^{s-2}\sum_{j=0}^{\frac{k+1}{r}-1}a_{i,j}g(x)^jx^i+\sum_{j=1}^{\frac{k+1}{r}-1}a_{s-1,j}g(x)^jx^{s-1}\\
+\sum_{i=s}^{r-1}\sum_{j=0}^{\frac{k+1}{r}-1}a_{i,j}g^j(x)x^{i-s}h_{A_m}(x).
\label{eq:zxc}
\end{multline}

The code is defined as the set of evaluations of $f_a(x), a\in \ff^k$.}
\end{cnstr}  

\begin{theorem}\label{thm:5.2}
The code given by Construction \ref{cnstr_n_r_linear} is an $(n,k,r)$ LRC code with minimum distance satisfying
    \begin{equation}\label{eq:dl}
    d\geq n-k-\Big\lceil \frac{k}{r}\Big\rceil +1.
    \end{equation}
\end{theorem}
Note that the designed minimum distance in \eqref{eq:dl} is at most one less than the maximum possible value.
\begin{IEEEproof}
Note that the encoding is linear and the encoding polynomial in \eqref{eq:zxc} is of degree at most
   \begin{multline*}
   \Big(\frac{k+1}{r}-1\Big)(r+1)+(r-1)\\=k+1-r+\frac{k+1}{r}-1+r-1=k+\Big\lceil \frac{k}{r}\Big\rceil -1.
   \end{multline*}
The bound \eqref{eq:dl} follows.
The locality property 
follows similarly to Construction \ref{cnstr_n_r}. Indeed, if the symbol $f_a(\alpha)$ for $\alpha\in A_m$ is to be recovered, 
we need to find a polynomial of degree at most $s-2$  from $s-1$ interpolation points.
\end{IEEEproof}

{\em Remark:}  
\cite[Cor.~10]{gopalan2011locality} 
shows that $(n,k,r)$ LRC codes whose distance meets the bound \eqref{eq:erer} do not exist whenever $r$ divides $k$ and 
    \begin{equation}\label{eq:div}
    0<n-\frac{k(r+1)}{r}<r+1.
    \end{equation}
(There is no contradiction with Construction \ref{cnstr3} which assumes in addition that $(r+1)|n$ since these 
two divisibility conditions imply that the inequalities in \eqref{eq:div} cannot be simultaneously satisfied.)
In the case that $r|k$ and these inequalities are satisfied, the best
bound is at least one less than \eqref{eq:erer}, which implies that the codes discussed in Theorem \ref{thm:5.2} are optimal for the
considered values of $r,k,$ and $n.$

\subsection{LRC codes as Redundant Residue Codes} \label{sect:RRC}
So far in this paper we have discussed the problem of recovering the lost symbol of the codeword by accessing a specific
subset of $r$ other symbols. We presented a construction of optimal LRC codes with this functionality and several of its 
modifications. 
Of course, in order to locally recover a lost symbol, all the $r$ other symbols must be accessible. 
Having in mind the distributed storage application, we argue that this may not always be the case, for instance, if
the symbols of the codeword are distributed across a network, and some nodes of the network become temporarily inaccessible.
For this reason, in this section we consider a general method of constructing 
$(n,k,r)$ LRC codes such that every symbol is contained in an MDS local code with \emph{arbitrary} parameters.

More formally, for an integer $t$ let $n_1,...,n_t$ and $k_1,...,k_t$ be two sequences of integers that satisfy 
    $$
    k\le \sum_{i}k_i, n=\sum_{i}n_i \text{ and } k_i\leq n_i \text{ for any } i.
    $$ 
We will construct a code such that its symbols can be partitioned into $t$ codes  $\cC_i$, and each $\cC_i$ is an $(n_i,k_i)$ MDS code. 
The idea of the construction in this section is similar to the description of 
Reed-Solomon codes as redundant residue codes \cite[Sect. 10.9]{mac91} which relies on the Chinese Remainder Theorem.

\vspace*{.1in}\noindent
{\bf Chinese Remainder Theorem: }{\em
Let $G_1(x),...,G_t(x)\in \ff[x]$ be pairwise coprime polynomials, then for any $t$ polynomials 
$M_1(x),\dots,M_t(x)\in\ff[x]$ there exists a unique polynomial $f(x)$ 
of degree less than $\sum_{i}\deg(G_i)$, such that
   $$
f(x) \equiv M_i(x) \,\text{\rm mod}\, G_i(x) \quad\text{ for all } i=1,\dots,t.
$$
}
\begin{cnstr} \label{residue}{\rm
Let $A\subset\ff,|A|=n$ be a subset of points, and let $\cA=\{A_1,...,A_t\}$ be a partition of $A$ such that $|A_i|=n_i, i=1,\dots,t$. 
Let $\Psi$ be an injective mapping 
     \begin{align*}
     \Psi:&\ff^k\rightarrow \cF_{k_1}[x]\times ...\times \cF_{k_t}[x]\\
          &a\mapsto (M_1(x),\dots,M_t(x)),
     \end{align*}
where $\cF_{k_i}[x]$ is the space of polynomials of degree less than $k_i,i=1,\dots,t$. 
Let $$G_i(x)=\prod_{a\in A_i}(x-a),i=1,\dots,t$$ be the annihilator polynomial of the subset $A_i.$
Clearly the polynomials $G_i(x)$ are pairwise coprime.

For a message vector $a\in \ff^k$ define the encoding polynomial $f_a(x)$  
 to be the unique polynomial of degree less than $n$ that satisfies  
          $$
    f_a(x) \equiv M_i(x) \,\text{\rm mod\,} G_i(x).
         $$
Finally, the code is defined as the image of the evaluation map \eqref{eq:code1} for the set of message vectors
$\ff^k.$}
\end{cnstr}

\begin{theorem} 
Construction \ref{residue} constructs an $(n,k)$ LRC code with $t$ disjoint local codes $\cC_i$, where each $\cC_i$ is an $(n_i,k_i)$ MDS code.
\end{theorem}

\begin{IEEEproof} 
Since each codeword is an evaluation at $n$ points of a polynomial of degree less than $n$, 
the weight of each nonzero codeword is at least one, 
and the code defined by the construction is indeed an injective mapping of $\ff^k$ to $\ff^n$. 

Consider the set $A_i,i=1,\dots,t$ in the partition and note that by the construction, there exists a polynomial $h$ such that 
    $$
f_a(x)=h(x)G_i(x)+M_i(x).
    $$
This implies that $f(\alpha)=M_i(\alpha)$ for any $\alpha$ in $A_i$. 
In other words, the restriction of the codeword $(f_a(\alpha),\alpha\in A)$ to the subset of locations 
corresponding to $A_i$ can be viewed as an evaluation of a polynomial of degree less than $k_i$ at $n_i$ points.
Therefore, the vectors $(f_a(\alpha),\alpha\in A_i)$ form 
an $(n_i,k_i)$ MDS code for all $i=1,\dots,t.$
\end{IEEEproof}

The distance of the code constructed using the method discussed here is at least $\min_{1\le i\le t}(n_i-k_i+1).$
It is easy to see that Construction \ref{cnstr1.1} and Construction \ref{cnstr3} are 
special cases of Construction \ref{residue}, where each local code is an $(r+1,r)$ MDS code. 
Note also that Construction \ref{residue} provides significant flexibility, allowing one to combine
arbitrary local MDS codes into an LRC code.

\subsection{$(r+\rho-1,r)$ Local MDS Codes}
The construction considered in this section is a special case of the general construction of the previous section in which
all the local codes have the same parameters. More specifically, we consider LRC codes in which the set of coordinates is partitioned
into several subsets of cardinality $r+\rho-1$ in which every local code is an $(r+\rho-1,r)$ MDS code, where $\rho
\ge 3.$
Under this definition, any symbol of the codeword is a function of any $r$ out of the $r+\rho-2$ symbols, increasing the
chances of successful recovery. Such codes will be called 
$(n,k,r,\rho)$ LRC codes, where $n$ is the block length and $k$ is the code dimension (here we confine ourselves to the case
of linear codes). Kamath et al. \cite{kamath2012codes}
  generalized the upper bound \eqref{eq:erer} to $(n,k,r,\rho)$ LRC codes, showing that the minimum distance $d$ satisfies
\begin{equation}
d\leq n-k+1-\Big(\Big\lceil \frac{k}{r}\Big\rceil-1\Big)(\rho-1).
\label{eq:23}
\end{equation}
As before, we will say that the LRC code is optimal if its minimum distance attains this bound with equality.

We assume that $n|(r+\rho-1)$ and $r|k,$ although the latter constraint is again unessential.
The following construction is described for the case of linear codes, generalizing Construction \ref{cnstr3}. 
It is also possible to extend the more general Construction \ref{cnstr1.1} to the case at hand, however we will not include the details.

\begin{cnstr}{\rm
Let  $\cA=\{A_1,\dots,A_m\}, m=n/(r+\rho-1)$ be a partition of the set $A\subset \ff, |A|=n,$
such that $|A_i|=r+\rho-1,1\le i\le m$.
Let $g\in \ff[x]$ be a polynomial of degree $r+\rho -1$ that is constant on each of the sets $A_i$. The polynomials $1,g,...,g^{m-1}$ span the algebra $\ff_{\cA}[x]$, see Proposition \ref{prop1} part (3).
For an information vector $a\in \ff^k$ define the encoding polynomial
\begin{equation}
f_a(x)=\sum_{\substack{i=0\\ i\,\text{\rm mod} (r+\rho-1)=0,1,...,r-1}}^{k-1+ (\frac{k}{r}-1)(\rho-1)}
a_ig(x)^{\lfloor \frac{i}{r+\rho-1}\rfloor} x^{i\,\text{\rm mod} (r+\rho-1)}.
\label{eq:564}
\end{equation}

The code is the image of $\ff^k$ under the evaluation map, see \eqref{eq:code1}.}
\label{cnstr2}
\end{cnstr}

We note that the polynomial $f_a(x)$ can be also represented in the form analogous to \eqref{eq:97979}.
Indeed, let $a=(a_0,...,a_{r-1})\in \ff^{k}$, where each $a_i=(a_{i,0},...,a_{i,\frac{k}{r}-1})$ is a vector of length $k/r$.  For $i=0,...,r-1$ define  
   $$
   f_i(x)=\sum_{j=0}^{\frac{k}{r}-1}a_{ij}g(x)^j,
   $$
then \eqref{eq:564} becomes 
\begin{equation*}
  f_a(x)=\sum_{i=0}^{r-1}f_i(x)x^i,
\end{equation*}
 
\begin{theorem}
Construction \ref{cnstr2} yields an optimal $(n,k,r,\rho)$ LRC code.
\end{theorem} 
\begin{IEEEproof}
Since the degree of the encoding polynomial satisfies $\deg(f_a)\le k-1+(\lceil \frac{k}{r}\rceil-1)(\rho-1)$ and the code is linear,
we conclude that the bound on the code distance in \eqref{eq:23} is achieved with equality. The local recoverability 
property follows similarly to Theorem \ref{thm1.1}. Indeed, suppose that the erased symbol is $f_a(\alpha)$ for some $\alpha$ in $A_i.$
The restriction of $f_a$ to the set $A_i$ is a polynomial of degree at most $r-1$. 
At the same time, $|A_i\backslash\{\alpha\}|=r+\rho-2$, so $f_a$ can be reconstructed from any $r$ of its values on the locations
in $A_i.$ The theorem is proved.
\end{IEEEproof}

\section{Conclusions}
In this paper we constructed codes that meet the Singleton-like bound \eqref{eq:erer} on the minimum distance $d$ for any value of
the locality parameter $r, 1<r<k$. The codes form a natural generalization of Reed-Solomon codes, which takes the locality
condition into account. We also extended the main construction to codes with multiple independent recovering sets so that 
the lost symbol can be corrected by accessing several different $r$-subsets of the codeword coordinates.

In regards to future research directions related to the code family studied here, we mention
correction of up to $\lfloor \frac{d-1}{2}\rfloor$ errors with these codes using algebraic decoding
algorithms, as well as generalizations of our constructions to algebraic geometric codes.

\appendix[Proof of Theorem \ref{thm:S}]
We will use the following theorem which is a slight modification of the well-known Tur{\'a}n theorem on the size of the maximal independent set in a graph.
\begin{theorem}
\label{turan}
Let $G$ be a directed graph on $n$ vertices, then there exists an induced directed acyclic subgraph of $G$ on at least 
$$\frac{n}{1+\frac1n{\sum_{i}d_i^{\text{out}}}}$$ vertices, 
where $d_i^{\text{out}}$ is the outgoing degree of vertex $i$.
\end{theorem}
\begin{IEEEproof} 
We follow the proof of the undirected version of this result that appears in \cite[pp.95-96]{Alon08}.
Choose uniformly a total ordering $\pi$ on the set of vertices $[n]$, and let $U\subseteq [n]$ be a subset of vertices defined as follows: A vertex $i$ belongs to $U$ iff for any outgoing edge from 
$i$ to some vertex $j$, $\pi(i)<\pi(j)$.
The induced subgraph of $G$ on $U$ is a directed acyclic graph, since if $i_1,...,i_m$ is a cycle where $i_j\in U$ then 
$$\pi(i_1)<\pi(i_2)<...<\pi(i_m)<\pi(i_1),$$ and we get a contradiction. Let $X=|U|$ be the size of $U$, and let $X_i$ be the indicator random variable for $i\in U$. Clearly $X=\sum_i X_i$ and for each $i$
   $$
   E(X_i)=P(i\in U)=\frac{d_i^{\text{out}}!}{(1+d_i^{\text{out}})!}=\frac{1}{1+d_i^{\text{out}}}.
   $$ 
Using the inequality between the arithmetic mean and the harmonic mean, we obtain
$$E(X)=\sum_{i}\frac{1}{1+d_i^{\text{out}}}\geq \frac{n}{1+\frac{\sum_id_i^{\text{out}}}{n}}.$$
Therefore there exists a specific ordering $\pi$ with 
$$|U|\ge \frac{n}{1+\frac{\sum_id_i^{\text{out}}}{n}}.$$
\end{IEEEproof}
\begin{IEEEproof} (of Theorem \ref{thm:S})
Consider a directed graph $G$ whose vertex set is the set of coordinates $[n]$ of $\cC$, 
and there is a directed edge from $i$ to $j$ iff $j\in I_i$. 
Since the code has locality $r$, the outgoing degree of each vertex is at most $r$, and by Theorem \ref{turan} $G$ contains an induced directed acyclic subgraph $G_U$ on the set of vertices $U$, where 
\begin{equation}
|U|\ge \frac{n}{r+1}.
\label{eq:wzwz}
\end{equation}
Let $i$ be a coordinate in $G_U$ without outgoing edges, then it is clear that coordinate $i$ is a function of the coordinates $[n]\backslash U $.  Continuing with this argument, consider the induced subgraph of $G$ on $U\backslash i$. Clearly it is also a directed acyclic graph. Let $i'$ be another coordinate without outgoing edges in $G_{U\backslash i}$, which means that coordinate $i'$ is a function of the coordinates $[n]\backslash U$. Iterating this argument, we conclude that any coordinate $i\in U$ is a function of the coordinates $[n]\backslash U$.

This means that we have found at least $|U|\geq \frac{n}{r+1}$ coordinates that are redundant. 
Therefore, the number of information coordinates $k$ is at most $rn/(r+1)$, as claimed. 

For the second part note that the minimum distance of the code can be defined as follows:
    \begin{equation}\label{eq:d}
    d=n-\max_{I\subseteq [n]}\{|I|: |\cC_I|<q^k\}.
    \end{equation}
{Consider a subset $U'\subseteq U$ of size $|U'|=\lfloor \frac{k-1}{r}\rfloor$. 
Such a subset exists because using \eqref{prop546} and \eqref{eq:wzwz} we have
  $$
  |U|\geq \frac{n}{r+1}\geq \frac{k}{r}\geq \Big\lfloor\frac{k-1}{r}\Big\rfloor.
  $$}
Clearly the induced subgraph of $G$ on $U'$ is a directed acyclic graph. Let $N$ be the set of coordinates in $[n]\backslash U'$ that have at least one incoming edge from a coordinate in $U'$. Note that 
$$|N|\leq r|U'|=r\Big\lfloor\frac{k-1}{r}\Big\rfloor \leq k-1,$$
and that each coordinate in $U'$ is a function of the coordinates in $N$.
Let $N'$ be a $(k-1)$-element set formed by the union of $N$ with arbitrary $k-1-|N|$ coordinates from the set $[n]\backslash(N\cup N')$.
Hence 
$$|C_{N'\cup U'}|=|C_{N'}|\leq q^{k-1},$$
and $|N'\cup U'|=k-1+\lfloor \frac{k-1}{r} \rfloor$. 
Then we conclude that 
    $$
\max_{I\subseteq [n]}\{|I|: |\cC_I|<q^k\}\geq  k-1+\floorenv{ \frac{k-1}{r}},
    $$
and, using \eqref{eq:d},
   $$
d\leq n-\Big(k-1+\floorenv{ \frac{k-1}{r} }\Big)=n-k-\ceilenv{ \frac{k}{r}}+2.
   $$
\end{IEEEproof}

\providecommand{\bysame}{\leavevmode\hbox to3em{\hrulefill}\thinspace}
\providecommand{\MR}{\relax\ifhmode\unskip\space\fi MR }
\providecommand{\MRhref}[2]{%
  \href{http://www.ams.org/mathscinet-getitem?mr=#1}{#2}
}
\providecommand{\href}[2]{#2}

\vspace*{.5in}
\noindent{\bf Itzhak Tamo} 
was born in Israel in 1981. He received a B.A. from the
Mathematics Department, B.Sc. and Ph.D. from the Electrical and Computer
Engineering Department, at Ben-Gurion University, Israel. 
His research interests include: Storage systems and devices, coding theory,
computer systems design, and fault tolerance

\vspace*{.5in}
\noindent{\bf Alexander Barg} received his Ph.D. degree in Electrical Engineering from the Institute for 
Information Transmission Problems of the Russian Academy of Sciences, Moscow, Russia. His 
research interests include coding and information theory, signal processing, and algebraic combinatorics.
\end{document}